\documentclass[11pt]{scrartcl} 
\usepackage[paper=letterpaper]{geometry}
\usepackage[T1]{fontenc} 
\usepackage[USenglish]{babel} 
\usepackage{graphicx} 
\usepackage{makeidx} 
\usepackage{remreset} 
\usepackage[nouppercase]{scrpage2} 
\usepackage{amsmath} 
\usepackage{amssymb} 
\usepackage[
thmmarks, 
amsmath, 
hyperref 
]{ntheorem} 
\usepackage{bm} 
\usepackage{bbm} 
\usepackage{enumitem} 
\usepackage[hang]{subfigure} 
\usepackage{wrapfig} 
\usepackage{tabularx} 
\usepackage{multirow}
\usepackage{dcolumn} 
\usepackage{booktabs} 
\usepackage{listings} 
\usepackage{psfrag} 
\usepackage{multicol}
\usepackage{longtable}
\usepackage{ltxtable}
\usepackage{supertabular, listliketab}
\usepackage[table]{xcolor}
\usepackage{lscape}
\usepackage[normal]{caption}
\usepackage{rotating}
\usepackage{color}
\usepackage{chicago}
\usepackage{algorithmic}
\usepackage{algorithm}
\usepackage{caption}
\usepackage{leading}

\usepackage{appendix}

\newcommand\mytoday{\number\year-\ifcase\month\or 01\or 02\or 03\or 04\or 05\or 06\or 07\or 08\or 09\or 10\or 11\or 12\fi-\ifcase\day\or 01\or 02\or 03\or 04\or 05\or 06\or 07\or 08\or 09\or 10\or 11\or 12\or 13\or 14\or 15\or 16\or 17\or 18\or 19\or 20\or 21\or 22\or 23\or 24\or 25\or 26\or 27\or 28\or 29\or 30\or 31\fi} 
\setkomafont{sectioning}{\normalcolor\bfseries} 
\clearscrheadfoot 
\automark[section]{subsection} 
\setkomafont{captionlabel}{\bfseries} 
\newcolumntype{d}[2]{D{.}{.}{#1.#2}} 
\theoremstyle{break} 
\theoremheaderfont{\bfseries}
\theorembodyfont{\itshape}
\theoremseparator{}

\theoremstyle{nonumberbreak} 
\theoremsymbol{$\Box$}

\newcommand{\btheta}{\mbox{\boldmath $\theta$}}
\newcommand{\bOmega}{\mbox{\boldmath $\Omega$}}
\newcommand{\bomega}{\mbox{\boldmath $\omega$}}
\newcommand{\boeta}{\mbox{\boldmath $\eta$}}
\newcommand{\btau}{\mbox{\boldmath $\tau$}}

\begin{document}


\leading{18pt}

\begin{center}
{ \Large \noindent { \bfseries Detecting regime switches in the dependence structure of \\ high dimensional financial data} }
\end{center}

\vspace{1cm} \begin{center}
\begin{tabular}{lcl}
{  \Large Jakob  ST\"OBER}                &  and \qquad \qquad  & { \Large Claudia CZADO}   \\
                                   &                    &                        \\
  Zentrum Mathematik &                    &Zentrum Mathematik\\
{\it  Technische Universit\"at M\"unchen}  &                   & {\it  Technische Universit\"at M\"unchen} \\
  Boltzmannstr. 3                  &                    &Boltzmannstr. 3 \\
 D-85747 Garching, Germany         &                    &D-85747 Garching, Germany \\
 E-mail: stoeber@ma.tum.de        &                    &  E-mail: cczado@ma.tum.de \\
 Tel.: +49 89 28917425        &                    &   Tel.: +49 89 28917428 \\
 Fax: +49 89 28917435          & &   Fax: +49 89 28917435
\end{tabular}
\end{center}

\vspace{4cm} \begin{center}
\begin{tabular}{lc}
{\bf  Corresponding author:}                &    \qquad \qquad      \\
Jakob  St\"ober                &    \qquad \qquad      \\
  Zentrum Mathematik &                    \\
Technische Universit\"at M\"unchen  &\\
  Boltzmannstr. 3                  &          \\
 D-85747 Garching, Germany         &            \\
 E-mail: stoeber@ma.tum.de        &              \\
 Tel.: +49 89 28917425        &                    \\
 Fax: +49 89 28917435          &
\end{tabular}
\end{center}

\newpage

\vspace{1cm} \noindent {\bf Abstract}\\
Misperceptions about extreme dependencies between different financial assets have been an important element of the recent financial crisis. This paper studies inhomogeneity in dependence structures using Markov switching regular vine copulas. These account for asymmetric dependencies and tail dependencies in high dimensional data. We develop methods for fast maximum likelihood as well as Bayesian inference.
Our algorithms are validated in simulations and applied to financial data. We find that regime switches are present in the dependence structure of various data sets and show that regime switching models could provide tools for the accurate description of inhomogeneity during times of crisis.

\vspace{1cm} \noindent   {   \bf Keywords:}
Copula, R-vine, financial returns, pair-copula
construction, Markov switching

\vspace{1cm} \noindent {\bf JEL classification:} C13, C51, C32

\section{Introduction}	

It is a well-known fact for univariate financial time series, that variances are not constant over time and tend to cluster together, as it is described in Engle's seminal paper \cite{engle1982}. In the multivariate setting, however, investigating whether also the dependence structure between different assets is varying over time is far more challenging. 

In this paper, we propose to combine the very flexible class of regular vine (R-vine) copulas with a hidden Markov structure, which accounts for the stylized facts observed in the dependence structure of multivariate financial return series and to investigate whether different dependence periods are present in given data.
It is observed for economic data that the dependence is asymmetric between negative and positive returns and that also tail dependencies exhibit asymmetries, i.e. differ between the upper and lower tail (\citeN{longin2001}, \citeN{ang2002b}).
While classical models for multivariate time series, which are based on multivariate normal or Student-t distributions, fail to explain these observations, R-vine copulas utilize the richness of the class of bivariate copulas to account for such types of dependence \shortcite{joe2009}.
These copula models, which are constructed based on a series of linked trees called R-vine, constitute a new class of multivariate dependence models which has been introduced by \citeANP{bedford2001} \citeyear{bedford2001,bedford2002} using and generalizing ideas of \citeN{joe1996}. Since R-vine distributions are built up hierarchically from bivariate copulas as building blocks, also inference can be performed exploiting that hierarchical structure as pioneered in the seminal paper of \shortciteN{aas2009}. 
Further, they allow to scale dependence modeling to large dimensions by using truncation techniques as introduced by \citeN{heinen2009} and \citeN{valdesogo2009}, and considered more general in \shortciteN{brechmann2011b}. The hierarchical structure is exploited for very fast yet asymptotically efficient sequential estimation \cite{haff2010}. Applications in dimensions as high as 52 \cite{brechmann2011} or 100 \cite{heinen2009b} have been considered using restricted subclasses of the general R-vine model we propose.

Being exposed to the challenge of high dimensionality in many applications, research in the area of multivariate dependence modeling is focussed on the case of time-homogeneous dependence structures. However, there are also promising approaches for allowing variations in the dependence structures over time. Hereby, one popular direction of research uses parametric dependence models in which the parameters constitute a function of time. For example, \citeN{hafner2010} and \citeN{almeida2011} model dependence parameters as a latent autoregressive stochastic process. Another popular direction combines multivariate dependence models with regime switching and in particular Markov switching (MS) models. Publications in this direction include  \shortciteN{pelletier2006}, \citeN{garcia2011}, \citeN{okimoto2008}, \shortciteN{markwart2009}, and in particular \shortciteN{cholette2009}, who first applied a restrictive vine distribution in the context of time varying dependence models.
A recent survey, comparing these different approaches with focus on the bivariate case, is given in \shortciteN{manner2011}.

In this paper, we will pick up the second approach and extend the applicability of regime switching dependence models to high dimensions using a general R-vine model, extending the initial approach of  \shortciteN{cholette2009}. Since high dimensionality and the presence of latent state variables make parameter estimation challenging and much of econometric literature on hidden Markov and MS models only marginally considers inference methods, our main concern will be to fill this gap.

In this respect, we give a thorough introduction to parameter inference for MS models, extending the existing procedures in several ways.
While existing models for regime switching dependence structures lack the possibility to be extended to high dimensions, we demonstrate how parameter inference for the MS R-vine model can be facilitated in almost arbitrary dimension by using a fast approximative Expectation - Maximization (EM) procedure in a Maximum Likelihood (ML) framework. 
The developed algorithm can also be used to obtain proper starting values to perform Bayesian parameter inference using MCMC techniques. We extend the Bayesian estimation procedure which has been developed by \citeN{min2010} for Student-t copulas on drawable (D-)vines to the case of general R-vine distributions with arbitrary bivariate copulas and include inference about the underlying MS model as it has been considered by \citeN{kim1998}. This Bayesian inference method enables us to assess the uncertainty in parameter estimates by considering Bayesian credible intervals (CIs). This has not been possible for MS dependence models, because of the computational burden necessary to achieve bootstrapped confidence intervals.

In order to demonstrate the applicability and performance of our procedures for parameter estimation, we perform a simulation study and investigate several applications to empirical data.
In this context, we will also show how model selection can be performed for time-varying dependence structures conducting a rolling window analysis and compare different models using the Bayesian deviance information criterion (DIC, \shortciteN{spiegelhalter2002}).

The remainder of this paper is structured as follows: Section \ref{sec_model} introduces the Markov switching regular vine copula model by first introducing R-vine distributions in Section \ref{sec_model_rvine} and then combining them with an underlying Markov structure in Section \ref{sec_model_markov}. Section \ref{sec_inference} then focusses on the problem of parameter estimation, discussing the stepwise procedure in Section \ref{sec_inference_stepwise} and Bayesian estimation in Section \ref{sec_inference_bayesian}. In Section \ref{sec_simstudy} we present the results of our simulation study before turning to empirical applications in Section \ref{sec_applications}. We consider three data sets: nine exchange rates in Section \ref{sec_applications_FX}, five Eurozone country stock indices in Section \ref{sec_applications_index}, and the returns of ten selected stocks in the German stock index DAX in Section \ref{sec_applications_dax}.
Section \ref{sec_conclusion} concludes our paper and gives an outlook to directions of future research.

\section{The Markov switching regular vine copula model}\label{sec_model}

\subsection{Regular vine distributions}\label{sec_model_rvine}
\begin{figure}[!ht]
\centering
\includegraphics[width=0.8\textwidth]{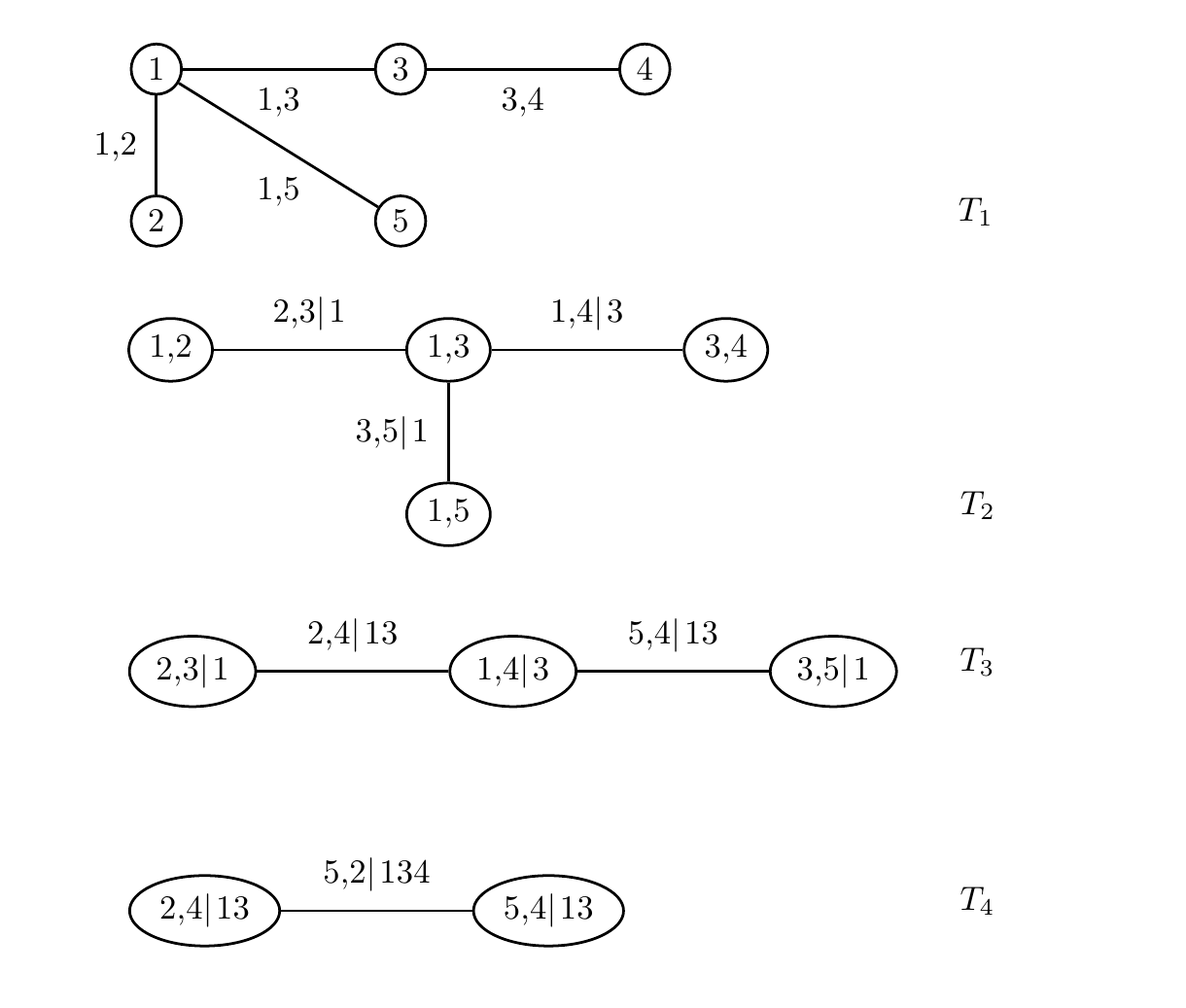}
\vspace{-.5cm}
\caption{An R-vine tree sequence in five dimensions with edge indices.}
\label{example_rvine}
\end{figure}

Regular vines as a graph theoretic tool for the construction of multivariate distributions have been introduced by \citeANP{bedford2001} \citeyear{bedford2001,bedford2002}. 
The R-vine itself is specified as a sequence of linked trees. It can be used to construct a multivariate distribution model by assigning to each edge a copula which corresponds to a bivariate conditional distribution, determined by the vine.
For this construction to be possible, the regular vine $\mathcal{V}$ on $d$ variables, which consists of a sequence of connected trees $T_1,\dots, T_{d-1}$, with nodes $N_i$ and edges $E_i$, $1\leq i \leq d-1$, needs to satisfy the following properties (\citeN{bedford2001}):
\begin{enumerate}
\item $T_1$ is a tree with nodes $N_1=\{1,\dots ,d\}$ and edges $E_1$. 
\item For $i \geq 2$, $T_i$ is a tree with nodes $N_i=E_{i-1}$ and edges $E_i$.
\item If two nodes in $T_{i+1}$ are joined by an edge, the corresponding edges in $T_i$ must share a common node. (\emph{proximity condition})
\end{enumerate}
There are two popular subclasses of R-vines which differ in the number of edges per node on each level. We call an R-vine
\begin{itemize}
\item \textbf{Canonical vine} (C-vine) if in each Tree $T_i$ there is one node which has edges with all $d-i$ other nodes.
\item \textbf{Drawable vine} (D-vine) if each node has edges with at most two other nodes.
\end{itemize}
A five-dimensional R-vine is shown in Figure \ref{example_rvine}. 
The notation we employ throughout our paper follows \citeN{czado2010}. To build up a statistical model using the R-vine, we associate to each edge $j(e),k(e)\vert D(e)$ in $E_i$, for $1 \leq i \leq d-1$, a bivariate copula density $c_{j(e),k(e)\vert D(e)}$, i.e. the copula associated to an edge $e$ is chosen to correspond to the copula of the conditional distribution indexed by $j(e),k(e)\vert D(e)$. We call $j(e)$ and $k(e)$ the {\it conditioned set} while $D(e)$ is the {\it conditioning set}. 
Let $\textbf{X}_{D(e)}$ denote the subvector of a vector $\textbf{X}$ determined by the set of indices $D(e)$ and $\textbf{X}_{-j}$ the subvetor of $\textbf{X}$ where the $j^{\text{th}}$ entry is deleted. For the definition of the  {\it regular vine distribution} we associate the bivariate copulas of the vine with bivariate conditional distributions. Let $\textbf{X}=(X_1,\dots,X_d)$ be a random vector with marginal densities $f_1,\dots,f_d$ respectively, and copula densities corresponding to the conditional distributions of $X_{j(e)}$ and $X_{k(e)}$ given $X_{D_e}$ equal to $c_{j(e),k(e)\vert D_e}$, then we call its distribution a regular vine distribution. In this case, the density is uniquely determined and given by 
\begin{gather}\label{rvine_density}
f_{1,\dots, d}(x_1,\dots,x_d)=\prod_{i=1}^d f_i(x_i)\cdot \prod_{i=1}^{d-1}\prod_{e \in E_i} c_{j(e),k(e)\vert D_e}(F(x_{j(e)}\vert \textbf{x}_{D_e}),(F(x_{k(e)}\vert \textbf{x}_{D_e}))
\end{gather}
as shown by \citeN{bedford2001}. If the marginal densities are uniform on $[0,1]$, we call the distribution in (\ref{rvine_density}) an R-vine copula. Given an R-vine $\mathcal{V}$, a set of corresponding parametric bivariate copulas $\textbf{B}$ and their parameter vector $\btheta$, we denote the R-vine copula density by $c(.\vert \mathcal{V},\textbf{B}, \btheta)$.

While also other iterative decompositions of a multivariate density into bivariate copulas and marginal densities are possible, R-vine distributions have the particularly appealing feature that the values for $F(x_{j(e)}\vert \textbf{x}_{D_e})$ and $F(x_{k(e)}\vert \textbf{x}_{D_e})$ appearing in Equation (\ref{rvine_density}) can be derived recursively without high dimensional integrations. To see how the tree structure is exploited for this, we refer the reader to Appendix \ref{appendix_stepwise} where it will become clear in the context of step-wise parameter estimation.

There are many possible R-vine structures in $d$ dimensions and the indices in (\ref{rvine_density}) do have a special order only in subclasses like D-vines or C-vines, as considered by \shortciteN{aas2009}. For this reason, we apply the approach of \citeN{morales-napoles2008}, \citeN{dissmann2010} and \shortciteN{dissmann2011} to store the indices in a $d \times d$ lower triangular matrix $M=(m_{ij} \vert i\geq j)$, where each row corresponds to a tree in the vine. In particular, given indices $m_{ii}$, $m_{ki}$ and $(m_{k+1,i},\dots,m_{di})$ in the matrix, there is an edge in the R-vine with $\{j(e),k(e)\}=\{m_{ii},m_{ki}\}$ and $D(e)=\{ m_{k+1,i},\dots,m_{di} \}$. For example, the matrix corresponding to the R-vine in Figure \ref{example_rvine} is 
\begin{gather*}
\begin{pmatrix}
5 & & & & \\
2 & 2 & & & \\
4 & 4 & 1 & & \\
3 & 3 & 4 & 3 & \\
1 & 1 & 3 & 4 & 4 \\
\end{pmatrix}.
\end{gather*}
The types of bivariate copulas in $\textbf{B}$ as well as their parameters $\btheta$ can conveniently be stored in matrices related to $M$, for details we refer to \citeN{dissmann2010}.

\subsection{Markov switching models}\label{sec_model_markov}

In this section, we introduce general MS models, as they have been established in statistics and econometrics by \citeN{hamilton1989}, focussing on the special case of multivariate dependence modeling. MS models constitute a special class of regime switching models, in which the process describing regime switches has a Markov structure. In essence, they assume that a hidden underlying process, which can be understood e.g. as the "state" of the world or the economy in financial applications, influences the development of a time series.

In particular, we will assume that the underlying process affects the dependence structure of a multivariate time series. The different possible dependencies can then be described by different R-vine copulas for each regime $k$, of which we assume the densities $c(.\vert   (\mathcal{V}, \textbf{B}, {\mbox{\boldmath $\theta$}})_k)$ to be given.
At each point in time, the present regime will determine the copula of the multivariate time series.
For this, let $(S_t)_{t=1,2,\dots}$ be a homogeneous discrete time Markov chain with states $\{1,\dots, p\}$. For simplicity, we assume it to be of first order such that it can be completely characterized by its transition matrix $P$ with elements $P(S_t=i\vert S_{t-1}=j)=P_{i,j}$. In applications where certain regressor variables influencing the development of the time series are known, our model could easily be extended to the case where the probabilities in this matrix change over time depending on these variables (c.f. \citeN{filardo1998}).

Given the Markov chain $(S_t)_{t=1,2,\dots}$, a MS R-vine copula model for a multivariate time series $(\textbf{u}_t=(u_{1t},\dots, u_{dt}), t=1,2,\dots)$ where $u_{it} \in [0,1] \ \forall i, t$ can be fully characterized by specifying conditional densities $$c(\textbf{u}_t\vert  (\mathcal{V}, \textbf{B}, {\mbox{\boldmath $\theta$}})_{1,\dots,p}, S_t) = \sum_{k=1}^p 1_{\{S_t=k\}} \cdot c(\textbf{u}_t\vert   (\mathcal{V}, \textbf{B}, {\mbox{\boldmath $\theta$}})_k).$$
The complete MS R-vine model is thus specified in terms of $p$ R-vine copula specifications and the matrix $P$ which contains the parameters of the underlying Markov chain. Considering inference in this context, we will always assume the R-vine structures $\mathcal{V}_k$ and corresponding sets of copulas $\textbf{B}_k$, $k=1,\dots, p$, to be given and thus suppress them in the following notation. The MS R-vine copula is then completely described by its parameters 
$$\btheta=(\btheta_{cop},\btheta_{MC})=((\btheta_1,\dots,\btheta_p),\btheta_{MC}),$$
where the subscript "cop" stands for copula parameters and "MC" for parameters making up the transition matrix $P$. 
Note that, although until now this is a pure copula model where no marginal time series structure is included, this introduces serial dependence. Given previous realizations it will be more or less likely that the hidden variable $S_t$ assumes a specific state. The individual marginal time series $(u_{i,t})_{t=1,2,\dots}$ however are i.i.d. uniform for $i=1,\dots,d$.

\section{Inference for Markov switching models}\label{sec_inference}

The first problem in developing inference methods for Markov Switching models is that we are faced with unobserved latent variables. In order to derive an expression for the full likelihood of a time series of observations $\tilde{\textbf{u}}_T = (\textbf{u}_1,\ldots,\textbf{u}_T)$, let us consider a decomposition of their joint density into conditional densities:
\begin{equation*}
\begin{split}
&f(\tilde{\textbf{u}}_T\vert \btheta) =f(\textbf{u}_1\vert \btheta) \cdot \prod_{t=2}^T f(\textbf{u}_t\vert \tilde{ \textbf{u}}_{t-1}, \btheta)\\  &\ \ \ \ \ \ \ \ = \left[\sum_{k=1}^p f(\textbf{u}_1\vert S_1=k, \btheta_k) P(S_1=i\vert \btheta_{MC}) \right] \cdot \prod_{t=2}^T \left[\sum_{k=1}^p   f(\textbf{u}_t\vert S_t=k, \btheta_k) \cdot P(S_t=k \vert\tilde{ \textbf{u}}_{t-1}, \btheta )\right],
\end{split}
\end{equation*}
where $\tilde{\textbf{u}}_t := (\textbf{u}_1,\ldots,\textbf{u}_t)$.
The unconditional probabilities $P(S_1=i)$ in this expression are known from the stationary distribution of the Markov chain, which we assume to exist. To obtain the probabilities
\begin{gather*}
\big(\bOmega_{t\vert t-1}(\btheta)\big)_k:= P(S_t=k \vert \tilde{ \textbf{u}}_{t-1}, \btheta)
\end{gather*}
we can apply the filter of \citeN{hamilton1989}. Assuming $\bOmega_{t-1\vert t-1}$ to be given, we calculate
\begin{equation*}
\begin{split}
& \bOmega_{t \vert t-1}(\btheta)= P \ \cdot \ \bOmega_{t-1 \vert t-1}(\btheta) \ \ \ \text{and}\\
&\bOmega_{t \vert t}(\btheta)=\frac{\bOmega_{t \vert t-1}(\btheta) \odot  (f(\textbf{u}_t \vert S_t=k,\tilde{ \textbf{u}}_{t-1},\btheta_k))_{k=1,\dots,p}}{\sum_{k=1}^p  \big(\bOmega_{t \vert t-1}(\btheta)\big)_k \odot  f(\textbf{u}_t \vert S_t=k,\tilde{ \textbf{u}}_{t-1},\btheta_k)}, \end{split}
\end{equation*}
and obtain all probabilities, which are required to evaluate the density, recursively. The operator $\odot$ denotes componentwise multiplication of two vectors. 

Because of the latent state variables and the resulting dependence between parameters, direct maximization of the likelihood for given $ (\mathcal{V}_k, \textbf{B}_k)$, $k=1,\dots,p$. is analytically not possible and numerically difficult. In the following, we discuss a frequentist and a Bayesian approach to make inference for this kind of model tractable.

\subsection{EM algorithm for MS models}\label{sec_inference_stepwise}
\citeN{hamilton1990} proposed to overcome the problems in maximum likelihood estimation for an MS model by using an Expectation - Maximization (EM) type algorithm. This algorithm iteratively determines parameter estimates $\btheta^l$, $l=1,2,\dots,$ which, under several regularity conditions, converge to the ML estimate for $l \rightarrow \infty$. In particular, it iterates the following steps: \begin{enumerate}
\item Expectation step: Obtain the conditional probabilities of the latent states $\tilde{\textbf{S}}_T=(S_1,\dots,S_T)$ given the current parameter vector $\btheta^l$, i.e. $P(S_t=s_t \vert \tilde{\textbf{u}}_T, \btheta^l)$.
\item Maximization step: Maximize a pseudo likelihood for $\btheta^{l+1}$, where the probability of being in a latent state $S_t=s_t$ is replaced by the probability from step 1.
\end{enumerate}

More precisely, let us define the expected pseudo log likelihood function for $\btheta^{l+1}$, given the observations $\tilde{\textbf{u}}_T$ and the current parameter estimate $\btheta^l$, as
\begin{equation}\label{em_cond_ex}
Q(\btheta^{l+1};  \tilde{ \textbf{u}}_T, \btheta^{l}) := \int_{\tilde{\textbf{S}}_T}\log\left(f(\tilde{ \textbf{u}}_T, \tilde{\textbf{S}}_T\vert \btheta^{l+1} )\right) f(\tilde{ \textbf{u}}_T, \tilde{\textbf{S}}_T\vert \btheta^{l} ),
\end{equation}
where $\int_{\tilde{\textbf{S}}_T}$ is short notation for summation over possible values of $\tilde{\textbf{S}}_T$, i.e.  \\$\int_{\tilde{\textbf{S}}_T} g(\tilde{\textbf{S}}_T) = \sum_{s_1=1}^n\ldots \sum_{s_t=1}^n g(S_1=s_1,\dots, S_T=s_T)$ for an arbitrary function g of $\tilde{\textbf{S}}_T$.
 
With this specification, step 2 can be rewritten as
\begin{itemize}
\item[2.] Maximize $Q(\btheta^{l+1};  \tilde{ \textbf{u}}_T, \btheta^{l})$ with respect to $\btheta^{l+1}$.
\end{itemize}

We will now investigate how these steps can be performed in the context of the MS R-vine copula model. 
The notation in this section is chosen to follow the notation of Kim and Nelson (1999), for a justification of the algorithm we refer to the original work of Hamilton (1990). 
 
For simplicity, let us further assume that the Markov chain $\tilde{\textbf{S}}_T$ has only two states, and denote $a:= P_{11}$, $b:=P_{22}$, then $\btheta_{MC}=(a, b)$ and $\btheta_{cop}=( \btheta_1, \btheta_2 )$. Using the Markov property of $\tilde{\textbf{S}}_T$, we can decompose the joint density of $(\tilde{\textbf{u}}_T,\tilde{\textbf{S}}_T)$ as
\begin{equation*}
\log\left(f(\tilde{ \textbf{u}}_T, \tilde{\textbf{S}}_T\vert \btheta )\right)=\sum_{t=1}^{T} \log\left(f(\textbf{u}_t\vert S_t, \btheta_{cop})\right) + \sum_{t=1}^{T} \log\left(P(S_t \vert S_{t-1}, \btheta_{MC})\right).
\end{equation*}
With this, we can simplify Expression (\ref{em_cond_ex}) using $\btheta^l=(\btheta^l_{cop}, \btheta^l_{MC})$ as
\begin{equation*}\label{em_simplify}
\begin{split}
& Q(\btheta^{l+1};  \tilde{ \textbf{u}}_T, \btheta^{l}) = \int_{\tilde{\textbf{S}}_T}\log\left(f(\tilde{ \textbf{u}}_T, \tilde{\textbf{S}}_T\vert \btheta^{l+1} )\right) f(\tilde{ \textbf{u}}_T, \tilde{\textbf{S}}_T\vert \btheta^{l} )\\ & \propto 
 \int_{\tilde{\textbf{S}}_T}\log\left(f(\tilde{ \textbf{u}}_T, \tilde{\textbf{S}}_T\vert \btheta^{l+1} )\right) P(\tilde{\textbf{S}}_T\vert \tilde{ \textbf{u}}_T, \btheta^{l} ) \ \ \ \ \text{(with respect to }\btheta^l) \\
&=\int_{\tilde{\textbf{S}}_T} \left[\sum_{t=1}^{T} \log\left(f(\textbf{u}_t\vert S_t, \btheta_{cop}^{l+1})\right) + \sum_{t=1}^{T} \log\left(P(S_t \vert S_{t-1}, \btheta_{MC}^{l+1})\right) \right] \cdot P(\tilde{\textbf{S}}_T\vert \tilde{ \textbf{u}}_T, \btheta^{l} ) \\
&= \sum_{t=1}^{T} \int_{\tilde{\textbf{S}}_T} \log\left(f(\textbf{u}_t\vert S_t, \btheta_{cop}^{l+1})\right) \cdot P(\tilde{\textbf{S}}_T\vert \tilde{ \textbf{u}}_T, \btheta^{l} )\\ &+\int_{\tilde{\textbf{S}}_T} \left[ \sum_{t=1}^{T} \log\left(P(S_t \vert S_{t-1}, \btheta_{MC}^{l+1})\right) \right] \cdot P(\tilde{\textbf{S}}_T\vert \tilde{ \textbf{u}}_T, \btheta^{l} ) = A + B, 
\end{split}
\end{equation*}
where
\begin{equation*}
A :=  \sum_{t=1}^{T} \sum_{S_t=1}^2  \log\left(f(\textbf{u}_t\vert S_t, \btheta_{cop}^{l+1})\right)\cdot P(\tilde{\textbf{S}}_t\vert \tilde{ \textbf{u}}_T, \btheta^{l} )= \sum_{t=1}^{T} \sum_{S_t=1}^2  \log\left(f(\textbf{u}_t\vert S_t, \btheta_{cop}^{l+1})\right)\cdot \big(\bOmega_{t\vert T}(\btheta^l)\big)_{S_t}
\end{equation*} and 
\begin{equation*}
B := \int_{\tilde{\textbf{S}}_T} \left[ \sum_{t=1}^{T} \log\left(P(S_t \vert S_{t-1}, \btheta_{MC}^{l+1})\right) \right] \cdot P(\tilde{\textbf{S}}_T\vert \tilde{ \textbf{u}}_T, \btheta^{l} ).
\end{equation*}

Hereby, the probability $\big(\bOmega_{t\vert T}(\btheta^l)\big)_{s_t}=P(S_t=s_t \vert \tilde{\textbf{u}}_T, \btheta^{l})$, to which we will refer as the "smoothed" probability of being in state $s_t$ at time $t$, can be determined from the output of the Hamilton filter by applying the following backward iterations, called Kim's smoothing algorithm.\begin{equation*}
 \left(\bOmega_{t \vert T}(\btheta^l)\right)_{s_t}=\left(\left(P^T \cdot \frac{ \bOmega_{t+1 \vert T}(\btheta^l)}{ \bOmega_{t+1 \vert t}(\btheta^l)} \right) \odot  \bOmega_{t \vert t}(\btheta^l)\right)_{s_t},
\end{equation*}
where the division is to be understood componentwise.

The second term $B$ can be simplified further using the Markov property:
\begin{equation*}
\begin{split}
B = & \int_{\tilde{\textbf{S}}_T} \left[ \sum_{t=1}^{T} \log\left(P(S_t \vert S_{t-1}, \btheta_{MC}^{l+1})\right) \right] \cdot P(\tilde{\textbf{S}}_T\vert \tilde{ \textbf{u}}_T, \btheta^{l} )=\\
& \sum_{t=1}^{T}\sum_{i,j=0}^1 \log\left(P(S_t=i \vert S_{t-1}=j, \btheta_{MC}^{l+1})\right) P(S_t=i, S_{t-1}=j \vert \tilde{\textbf{u}}_T, \btheta^{l}) = \\
& \sum_{t=1}^{T}\big[(\log(a)  P(S_t=0, S_{t-1}=0 \vert \tilde{\textbf{u}}_T, \btheta^{l}) + \log(1-a)  P(S_t=1, S_{t-1}=0 \vert \tilde{\textbf{u}}_T, \btheta^{l}) + \\
&  + \log(b) P(S_t=1, S_{t-1}=1 \vert \tilde{\textbf{u}}_T, \btheta^{l}) + \log(1-b)  P(S_t=0, S_{t-1}=1 \vert \tilde{\textbf{u}}_T, \btheta^{l}) )\big].
\end{split}
\end{equation*}
This has now to be maximized with respect to $a$ and $b$. Taking the derivative with respect to $a$ and $b$ respectively, we obtain the step $l+1$ estimates for $a$ and $b$:
\begin{equation*}
\begin{split}
a^{l+1}&=\frac{\sum_{t=1}^T P(S_t=0, S_{t-1}=0 \vert \tilde{\textbf{u}}_T, \btheta^{l})}{\sum_{t=1}^T P(S_{t-1}=0 \vert \tilde{\textbf{u}}_T, \btheta^{l})}\\
b^{l+1}&=\frac{\sum_{t=1}^T P(S_t=1, S_{t-1}=1 \vert \tilde{\textbf{u}}_T, \btheta^{l})}{\sum_{t=1}^T P(S_{t-1}=1 \vert \tilde{\textbf{u}}_T, \btheta^{l})}
\end{split}
\end{equation*} 

In contrast to the model originally considered by Hamilton where all maximization steps could be performed analytically, this is not possible for the maximization of $A$ with respect to the copula parameters in our case. This means that, while the transition probabilities can be obtained directly, the second part of the maximization step has to be performed using numerical optimization methods.
Since a $d$-dimensional R-vine copula specification, in which each pair copula has $k$ parameters, contains $d(d-1)/2\cdot k$ parameters, this is computationally still very challenging. To circumvent this problem, we can exchange the joint maximization with the stepwise procedure of Algorithm \ref{stepwise_estimates} (Appendix \ref{appendix_stepwise}).

We call this the {\it stepwise EM-Algorithm}. Since treewise estimation of copula parameters is asymptotically consistent, this constitutes a close approximation to the "proper" EM-Algorithm.
While there are theoretical results on the convergence of the EM-Algorithm \cite{wu1983}, we loose these properties with our approximation. All limit theorems however do rely on proper maximization at each step of the algorithm. This is almost impossible to guarantee in our case where we are faced with high dimensional optimization problems and have to rely on numerical techniques. Therefore, also an implementation of the "proper" EM-Algorithm has to be considered an approximation, which, apart from the asymptotic consistency of the step-wise estimator further justifies the use of a stepwise procedure.
While all existing models for time varying dependence structures in high dimensions suffer from the computational burden for numerical estimation, we do only need to maximize the likelihoods of bivariate copulas in this tree-wise procedure, which reduces computation time and avoids the curse of dimensionality.

The estimate obtained by iterating the EM steps until convergence will be denoted by
$$\hat \btheta^{EM}=\left( \hat \btheta^{EM}_{cop} = (\hat \btheta_1^{EM},\dots,\hat \btheta_k^{EM}), \hat \btheta_{MC}^{EM} \right).$$

\subsection{Gibbs sampling for Markov switching models}\label{sec_inference_bayesian}
Having derived an approximative ML procedure for our MS models, we will now consider Bayesian estimation methods, which will enable us to quantify the uncertainty in parameter estimates.
In particular, CIs and posterior standard deviations are determined naturally while the uncertainty in ML parameter estimates is very hard to assess in this context.
Bayesian estimation for MS models has originally been considered by \citeN{albert1993}. Building on their ideas, the Gibbs sampler which we develop in this section consists of updates for the copula parameters, the Markov chain parameters and the latent state vector, respectively.
To derive how these update steps are to be performed, let us first neglect the latent Markov part of the model specification and focus on the parameters $\btheta_{cop}$ of R-vine copulas.

\subsubsection*{Update of copula parameters}

In order to complete the model specification in a Bayesian framework, we first have to specify prior distribution for each component of $\btheta_{cop}$.
Following \citeN{min2010}, we assume non informative priors for all copula parameters in the model. For bivariate copula families considered where the parameter range is not compact, we restrict its support to some finite interval to avoid numerical instabilities for very small or large parameter values. 
If for all bivariate copulas there is a one-to-one correspondence between parameter values and Kendall's $\tau$ given in closed form, the approach of \citeN{hofmann2010} with uniform priors for $\tau$ can be considered as an alternative. Furthermore, we can use a uniform prior for the correlation matrix of the model if all bivariate building blocks are Gaussian or Student copulas, c.f. \shortciteN{lewandowski2009}.

Since the posterior distribution is analytically not tractable and in particular expressions for the conditional posterior distributions are not available, we utilize the Metropolis-Hastings (MH) algorithm. As for the prior distribution, there are several possible choices for proposal distributions in this case. \citeN{min2010} use a modification of standard random walk proposals where the normal distribution is truncated to the support of parameters, proposal variances are tuned to achieve suitable acceptance rates. This leads to poor acceptance rates in some cases with strong dependencies and to high autocorrelations in general. To overcome these problems, we consider a two point mixture of a random walk proposal with an independent normal distribution at the mode for each parameter. The modes of the individual parameters as well as their standard errors are approximated using the bivariate hessian from the stepwise estimation procedure for R-vines and both distributions are assigned a weight of 0.5. The methods of using independence proposals centered around the mode has been proposed by \citeN{gamerman2006} and it has been applied in a context similar to ours by \shortciteN{czado2010b}. 
While there are parameter constellations where pure random walk proposals are more favorable than independence proposals and vice versa, simulation studies showed that the chosen mixture distribution works well for all settings.
Given these considerations Step 3 of the Gibbs sampler can be performed using MH within Gibbs by sampling $\btheta_i$ for $i=1,\dots,p$ from $$f(\btheta_i \lvert  \textbf{u}_{\{t \in \{1,...T\} \vert S_t=i\}}).$$

\subsubsection*{Update of Markov chain parameters}

For the second step of the Gibbs sampler, we will assume independent Dirichlet distributions as prior distributions for the columns of the transition matrix $P$, i.e. $(P_{i,j})_{i=1,\dots,n} \sim \text{Dirichlet}\big( (\alpha_{i,j})_{i=1,\dots,n} \big)$. The likelihood function for the entries of $P$ given a realization of the state vector $\tilde{\textbf{S}}_T$ has the form of a multinomial distribution
\begin{gather*} l(P\vert \tilde{\textbf{S}}_T)= \prod_{j=1}^p\prod_{i=1}^p p_{ij}^{n_{ij}},
\end{gather*}
where $n_{ij}$ denotes the number of transitions from state $j$ to state $i$ in $\tilde{\textbf{S}}_T$. Since the Dirichlet and the multinomial distribution are conjugate distributions (see \shortciteN{kotz2000}), also the conditional posterior distributions are Dirichlet distributions with parameters $\alpha^{posterior}_{i,j}=\alpha_{i,j}+n_{i,j}$. From these we can sample directly.

\subsubsection*{Update of the latent state vector}

For the first step, we follow the approach by \citeN{kim1998}, who assume independent non informative priors for the latent states and draw $\tilde{\textbf{S}}_T$ as a block.
To do so, we decompose 
\begin{gather*}
P(\tilde{\textbf{S}}_T\vert \tilde{ \textbf{u}}_T) = P(S_T\vert \tilde{ \textbf{u}}_T) \cdot \prod_{i=1}^{T-1} P(S_t \vert S_{t+1},\dots S_T, \tilde{\textbf{u}}_T) = P(S_T\vert \tilde{ \textbf{u}}_T) \cdot \prod_{i=1}^{T-1} P(S_t \vert S_{t+1}, \tilde{\textbf{u}}_T),
\end{gather*}
which allows us to generate $S_T$ from $P(S_T \vert  \tilde{\textbf{u}}_T)$ and $S_t$ for $t \in \{T-1,\dots,1\}$ from 
\begin{gather*}
P(S_t \vert \tilde{ \textbf{u}}_t,S_{t+1}) \propto P(S_{t+1} \vert S_t) P(S_t \vert \tilde{ \textbf{u}}_t), 
\end{gather*}
where $P(S_t \vert \tilde{ \textbf{u}}_t)=\bOmega_{t\vert t}(\btheta)$ can again be determined using the Hamilton filter.
This corresponds to sampling $\tilde{\textbf{S}}_T$ from $f(\tilde{\textbf{S}}_T\vert \tilde{ \textbf{u}}_T, \btheta_{cop}, \btheta_{MC}).$
More formally, the three steps of the Gibbs sampler from above can now be reexpressed as follows: 
\begin{algorithm}[H]
\caption{MCMC sampling for MS R-vine copulas.}
 \label{gibbs_sampler}
\begin{algorithmic}
\STATE Sample $\tilde{\textbf{S}}_T$ from $f(\tilde{\textbf{S}}_T\vert \tilde{ \textbf{u}}_T, \btheta)$ using Gibbs sampling.\\
\STATE Sample $P$ from  $f(P \vert \tilde{\textbf{S}}_T)$ using Gibbs sampling. \\
\STATE Sample $(\btheta_i \lvert  \textbf{u}_{\{t \in \{1,...T\} \vert S_t=i\}})$, for $i=1,\dots,n$, using MH within Gibbs sampling. \\
\end{algorithmic}
\end{algorithm}
Iterating through these steps will yield 
$$(\btheta^{r,MCMC},\tilde{\textbf{S}}_T^{r,MCMC})=\left( \left( (\btheta_1^{r,MCMC},\dots,\btheta_n^{r,MCMC}), \btheta_{MC}^{r,MCMC} \right), \tilde{\textbf{S}}_T^{r,MCMC} \right),$$
for $r=1,\dots,R$, where $R$ is the number of sampled realizations from the posterior distribution.

\section{Simulation study}\label{sec_simstudy}

Having derived all necessary components of our posterior sampling algorithm, this section sums up the results of a simulation study which has been performed in order to demonstrate the ability of the developed Bayesian inference procedure to capture the true model in simulated data.

We consider two regimes in four dimensions: A multivariate Gaussian copula, modeled as a D-vine copula where all bivariate copulas are Gaussian, and a C-vine copula, where all bivariate copulas are Gumbel or rotated Gumbel copulas; these regimes are kept fixed. We further consider 6 scenarios with different model parameters, summarized in Table \ref{sim_scenarios} in terms of the corresponding values for conditional Kendall's $\tau$. In all scenarios we set the Markov parameters to $a=0.95$ and $b=0.9$, the corresponding prior distributions are chosen to be non informative. 

\begin{table}[H]
\begin{center}
{\small
\begin{tabular}{  c c  c c c c c  }
& & & \multicolumn{4}{c}{coverage probability} \\
& \multicolumn{2}{c}{conditional Kendall's $\tau$} & \multicolumn{2}{c}{90\% CI}  &  \multicolumn{2}{c}{95\% CI} \\
\hline
& Gumbel regime & Gauss regime &  symm. &  HPD &  symm. & HPD \\
\hline
\multirow{3}{*}{Scenario 1} &$\tau_{43\vert 21}=0.4$ &  $\tau_{41\vert 23}=0.4$ &\multirow{3}{*}{91.7\%} & \multirow{3}{*}{91.7\%}  & \multirow{3}{*}{94.2\%} & \multirow{3}{*}{94.2\%}  \\
&  $\tau_{42\vert 1}=0.6$, $\tau_{32\vert 1}=0.6$  &$\tau_{42\vert 3}=0.6$,  $\tau_{31\vert 2}=0.6$& & & & \\
 &$\tau_{41}=0.8$, $\tau_{31}=0.8$, $\tau_{21}=0.8$  &  $\tau_{43}=0.8$, $\tau_{32}=0.8$, $\tau_{21}=0.8$ & & & &\\
\hline

\multirow{3}{*}{Scenario 2} & $\tau_{43\vert 21}=0.4$ l  &  $\tau_{41\vert 23}=0.1$ &\multirow{3}{*}{89.2\%} & \multirow{3}{*}{89.2\%}  & \multirow{3}{*}{91.7\%} & \multirow{3}{*}{91.7\%} \\
&  $\tau_{42\vert 1}=0.6$, $\tau_{32\vert 1}=0.6$  &$\tau_{42\vert 3}=0.2$,  $\tau_{31\vert 2}=0.2$ & & & & \\
& $\tau_{41}=0.8$, $\tau_{31}=0.8$, $\tau_{21}=0.8$  &  $\tau_{43}=0.3$, $\tau_{32}=0.3$, $\tau_{21}=0.3$ & & & & \\
\hline
\multirow{3}{*}{Scenario 3} & $\tau_{43\vert 21}=0.1$ &  $\tau_{41\vert 23}=0.4$&\multirow{3}{*}{85\%} & \multirow{3}{*}{84\%}  & \multirow{3}{*}{91.7\%} & \multirow{3}{*}{93.4\%}  \\
 & $\tau_{42\vert 1}=0.2$, $\tau_{32\vert 1}=0.2$ &$\tau_{42\vert 3}=0.6$,  $\tau_{31\vert 2}=0.6$&&&& \\
& $\tau_{41}=0.3$, $\tau_{31}=0.3$, $\tau_{21}=0.3$   & $\tau_{43}=0.8$, $\tau_{32}=0.8$, $\tau_{21}=0.8$&&&& \\
\hline
\multirow{3}{*}{Scenario 4} & $\tau_{43\vert 21}=0.1$   &  $\tau_{41\vert 23}=0.1$ &\multirow{3}{*}{75\%} & \multirow{3}{*}{75\%}  & \multirow{3}{*}{81.7\%} & \multirow{3}{*}{92.5\%}  \\
 & $\tau_{42\vert 1}=0.2$, $\tau_{32\vert 1}=0.2$ &$\tau_{42\vert 3}=0.2$,  $\tau_{31\vert 2}=0.2$ &&&& \\
& $\tau_{41}=0.3$, $\tau_{31}=0.3$, $\tau_{21}=0.3$  & $\tau_{43}=0.3$, $\tau_{32}=0.3$, $\tau_{21}=0.3$ &&&& \\
\hline
\multirow{3}{*}{Scenario 5} & $\tau_{43\vert 21}=0.3$ &  $\tau_{41\vert 23}=0.3$ &\multirow{3}{*}{85.0\%} & \multirow{3}{*}{85.0\%}  & \multirow{3}{*}{92.8\%} & \multirow{3}{*}{92.8\%} \\
&  $\tau_{42\vert 1}=0.5$, $\tau_{32\vert 1}=0.3$ & $\tau_{42\vert 3}=0.5$,  $\tau_{31\vert 2}=0.3$ &&&& \\
& $\tau_{41}=0.7$, $\tau_{31}=0.5$ $\tau_{21}=0.3$  &  $\tau_{43}=0.7$, $\tau_{32}=0.5$, $\tau_{21}=0.3$ &&&&\\
\end{tabular}}
\vspace{-.2cm}
\caption{Simulation scenarios investigated and empirical coverage probabilities based on 120 data sets from each scenario.}
\label{sim_scenarios}
\end{center}
\end{table}

From each scenario, we simulate a time series with 800 four dimensional observations. As an input for the MCMC algorithm, we need the R-vine tree structure of the model, the corresponding set of bivariate copula families, and starting values for the parameters. Keeping the (true) structure and copula families we used for simulations, we obtain a posterior estimate for the parameters as follows:\begin{enumerate}
\item Starting values for the EM algorithm: Fit the copula for each regime to the whole data set using the stepwise estimation procedure, and cluster the observations according to their likelihood values. Refit the copula to the 400 observations which have the highest log likelihood. Set the MC parameters to $a=b=0.9$, since persistent regimes are expected.
\item Starting values for MCMC: Iterate the stepwise EM algorithm until convergence.
\item Obtain 1000 independent samples from the posterior distribution of the parameters.
The Gibbs sampler is started at the values obtained in Step 3, a burn-in period is discarded and the chain is sub-sampled according to the effective sample size.
\end{enumerate}

From the obtained samples, we estimate 90\% and 95\% symmetric and highest posterior density (HPD) CIs for the copula parameters and check whether all true copula parameters lie within these intervals. The HPD intervals are calculated using the "coda" package for R, while $\alpha$\% - symmetric intervals are determined by the empirical $\frac{\alpha}{2}$\% and $1-\frac{\alpha}{2}$\% quantiles. The procedure was repeated 120 times for each scenario with results reported in Table \ref{sim_scenarios}.
Relative bias and mean squared error (MSE) of two selected scenarios are displayed in Appendix \ref{appendix_simstudy}.
Since the parameters are not independent, the combination of CIs for single parameters does not necessarily yield a credible region for the same level. This, in addition to the fact that Bayesian CIs in a simulation study are hard to interpret in terms of their frequentist coverage, makes a statistical analysis of the results difficult.

But clearly the number of exceptions for each scenario, except Scenario 4, lies within the range of what we would expect, namely about 90\% (95\%) frequentist coverage. Scenario 4 corresponds to low dependence in both regimes, the higher number of exceptions in this case is due to identification problems.
We conclude, that with clearly distinguishable regimes the outlined procedure is able to identify the true model.


\section{Applications}\label{sec_applications}

In this section, we apply the MS model described in Section \ref{sec_model} together with the estimation procedures of Section \ref{sec_inference} to analyze three financial data sets. 
Since the focus of this paper is on modeling dependence structures of multivariate data, we apply a two step estimation approach as suggested by \citeN{joe1996b}. In the first step, appropriate parametric models for the marginal time series are fitted separately and used to transform the standardized residuals to approximately uniform margins. To this transformed data, we apply our copula model in the second step. While joint estimation of marginal and copula models as in \citeN{hofmann2010} is more effective and allows to take the estimation error in the models for the marginals  into consideration, it is also computationally more challenging. Given the size of our data sets we choose the two step procedures. Note that the marginal time series structure we impose does not hurt the assumptions for our copula model, since the standardized residuals form approximately an i.i.d. sample.

This proceeding implies that all statements regarding the dependence properties of investigated data do not refer to the dependence structure among the univariate marginal time series themselves. Instead, we describe the dependencies of their transformed residuals obtained in the two-step approach. While the marginal models account for time-varying conditional variances and autocorrelations in given data, the dependence properties between individual variables are uniquely determined by the dependence properties of their residuals. This means that, although our approach does not yield direct conclusions for observed variables, qualitative results and interpretations can always be transferred from the level of residuals to the level of observations.

Before Bayesian or frequentist parameter inference for the MS R-vine model can be conducted, appropriate R-vine structures and sets of bivariate copulas for each regime need to be selected in a preanalysis. To do so, we apply the heuristic model selection techniques as outlined in \citeN{dissmann2010}, \shortciteN{dissmann2011} and \shortciteN{brechmann2011b}. These are based on the following steps:
\begin{enumerate}
\item For each pair of variables estimate the corresponding value of Kendall's $\tau$ from the copula data set.
\item Create a fully connected graph which consists of the marginal variables as nodes and where an edge is added between every pair of variables.
\item Associate to each edge the absolute value of the corresponding Kendall's $\tau$ as a weight.
\item Determine the maximum spanning tree (MST), i.e. find a tree which maximizes the sum of edge weights using for example the algorithm of \citeN{prim}.
\item For each edge in the resulting tree fit a parametric bivariate copula from a catalogue of bivariate copula familis and estimate its parameters by ML. The specific family is then chosen using the Akaike information criterion (AIC).
\item For each pair of edges which share a common node apply the probability integral transformation of the conditional cdf given the common node based on the copula parameters estimated in step 4 to the corresponding copula data. This respects the proximity condition and provides pseudo observations for the next tree.
\item Proceed with the pseudo observations as in steps 1 to 6 until all trees together with their copula types and parameters are determined.
\end{enumerate}
In all our applications we assume the presence of two regimes. Unless mentioned otherwise, the copula families we will consider in step 5 are the Gauss copula and the Gumbel copula. Since the Gumbel copula is not invariant with respect to rotations, we consider its standard form and rotations by $90^\circ$, $180^\circ$ and $270^\circ$, respectively. For all models studied, we run the MCMC for around 20000 internal iterations discarding the first 1000 as burn-in, and keep every fifth observation to reduce autocorrelations. For estimating quantiles of the posterior distribution, we further thin the output according to what \shortciteN{kass1998} call the "effective sample size" (c.f. \citeN{carlin2009}). For stability reasons, we use this as a proxy for thinning with respect to the autocorrelation function itself. After this, we end up with ca. 1000 approximately i.i.d. samples.

\subsection{Exchange rates}\label{sec_applications_FX}
The first data set taken into consideration consists of 9 exchange rates against the US dollar, namely the Euro, British pound, Canadian dollar, Australian dollar, Brazilian real, Japanese yen, Chinese yuan, Swiss franc and Indian rupee. The observed time period is from July 22, 2005 to July 17, 2009, resulting in 1007 daily observations. The modeling of the one dimensional margins with appropriate ARMA-GARCH models and the transformation to copula data has been performed by \shortciteN{CzadoSchepsmeierMin2011}. In total, we consider 6 models, which will be defined as we proceed, their defining trees and the allowed copula families are listed in Table \ref{model_table}.

\begin{table}[H]
\begin{center}
{\small
\begin{tabular}{l c c c c c c }
\vspace{-.2cm}
& \ Regime 1\ && \ Regime 2 \  & \ Copulas\  &&\ Copulas \ \\
& (no crisis) && (crisis) & Regime 1&& Regime 2\\
\hline
Model (1) & $\mathcal{V}_{11}$ & $=$ & $\mathcal{V}_{12}=\mathcal{V}_1$ & mixed & $\neq$ &mixed \\
\hline
Model (2a) & $\mathcal{V}_{21}=\mathcal{V}_1$ & $\neq$ & $\mathcal{V}_{22}=\mathcal{V}_2$ & N & $\neq$ & SG \\
Model (2$\text{a}^\star$) & $\mathcal{V}_{21}=\mathcal{V}_1$ & $\neq$ & $\mathcal{V}_{22}=\mathcal{V}_2$ & N & $\neq$ & SG, N \\
Model (2b) & $\mathcal{V}_{21}=\mathcal{V}_1$ & $\neq$ & $\mathcal{V}_{22}=\mathcal{V}_2$ & N & $\neq$ & G \\
Model (2c) & $\mathcal{V}_{21}=\mathcal{V}_1$ & $\neq$ & $\mathcal{V}_{22}=\mathcal{V}_2$ & N & $\neq$ & Student-t \\
\hline
Model (3) & $\mathcal{V}_{31}=\mathcal{V}_1$ & $\neq$ & $\mathcal{V}_{32}=\mathcal{V}_3$ & mixed & $\neq$ & mixed \\
\end{tabular}
}
\vspace{-.3cm}
\caption{R-vine models considered for the exchange rate data.}
\label{model_table}
\end{center}
\end{table}

\subsubsection{R-Vine with switching parameters}\label{ex_Rvine}
As a first model for the exchange rate data set, we consider an R-vine with only MS parameters. To do so, we fit an R-vine with corresponding bivariate copulas to the data using the outlined procedure allowing for Gaussian (n), Gumbel (G), survival Gumbel (rotation by $180^\circ$, abb. as SG), and $90^\circ$ / $270^\circ$ rotated Gumbel (G90 / G270). As the estimated parameters for the bivariate copulas corresponding to the higher trees indicate conditional independencies, we truncate the R-vine copula after the second tree, i.e. we associate all edges on higher trees with independence copulas. The R-vine copula structure resulting from this procedure is tabled in Appendix \ref{appendix_rvine_structure}. We call this model Model (1).

Figure \ref{ex_rvine_switch} shows the probability $P(S_t=2 \vert \tilde{\textbf{u}}_T, \hat \btheta^{EM})$ that the hidden state variable $S_t$ indicates the presence of Regime 2 plotted against time. While Regime 1 is predominant until around February 2007, Regime 2 becomes more important during the later times of the financial crisis. 

\begin{minipage}[c]{\textwidth}
\begin{minipage}[c]{\textwidth}
\includegraphics[width=0.95\textwidth]{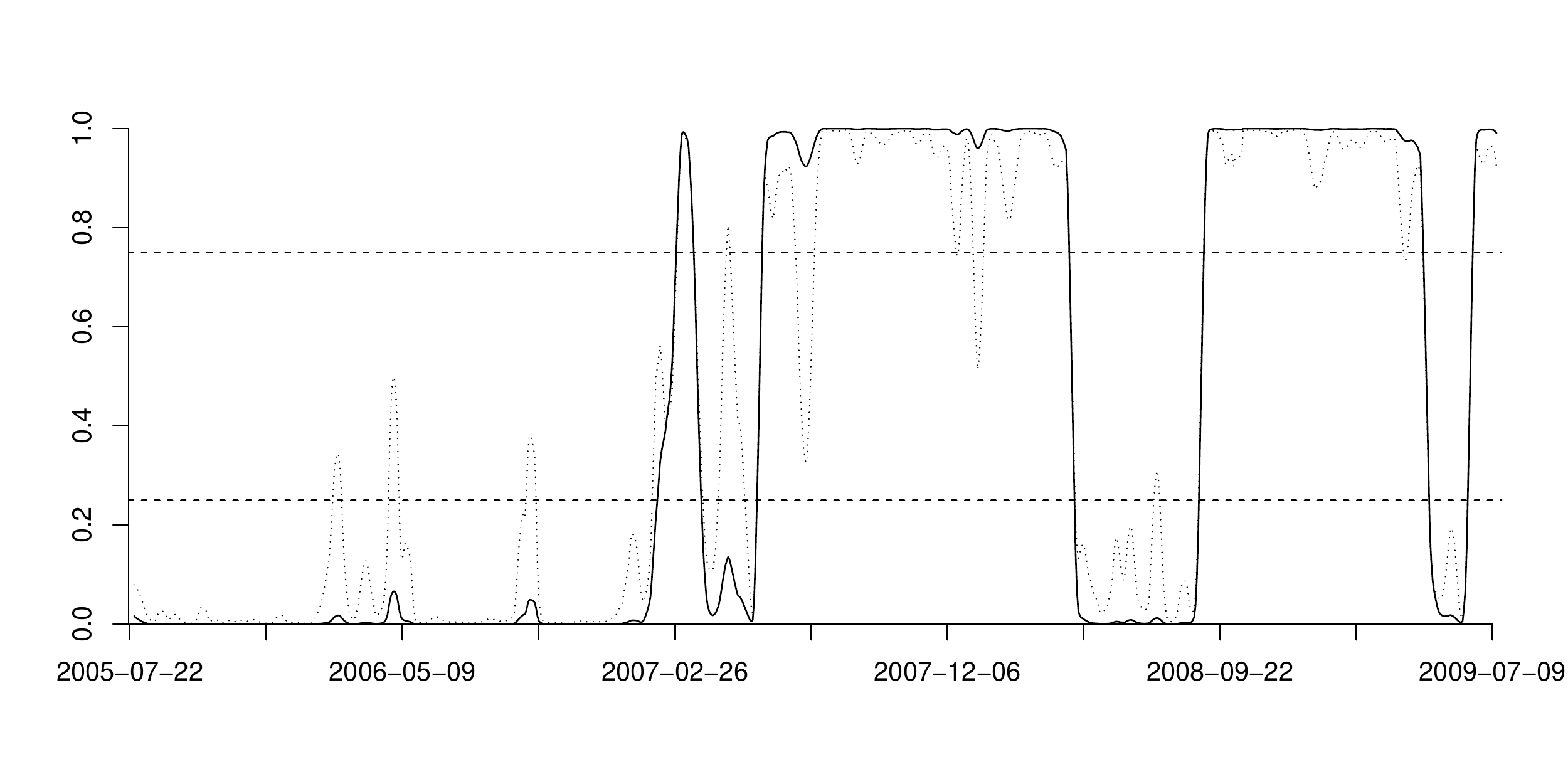}
\end{minipage}
\vspace{-1cm}
\captionof{figure}{Estimated probabilities over time of being in state 2 for Model (1). (Solid: EM estimates smoothed by an MA(7) filter, Dotted: Bayesian posterior estimates)}
\vspace{.3cm}
\label{ex_rvine_switch}
\end{minipage}

Analyzing the parameter estimates  $\hat \btheta^{EM}_1$ and $\hat \btheta^{EM}_2$ (see Table \ref{param_model1} of Appendix \ref{appendix_rvine_structure}) for the two regimes, we find that Regime 1 has stronger dependencies on the first tree, whereas Regime 2 has stronger dependencies on the second tree. In particular, Regime 2 exhibits stronger conditional negative dependencies reflected by rotated Gumbel copulas, thus creating a more asymmetric dependence structure.

In order to apply our Bayesian estimation procedure, we need to distinguish both regimes to avoid model identification problems. For a detailed consideration of this issue we refer the reader to \citeN{fruehwirth2001}. Using our observations with regard to the strength of dependence in the two regimes identified by the EM-Algorithm, we define Regime 1 to correspond to weaker dependence on the second tree and Regime 2 to correspond to stronger dependence on the second tree, compared by the sum of absolute values of Kendall's $\tau$ corresponding to parameters  $\btheta^{r,MCMC}_1$ and $\btheta^{r,MCMC}_2$.
The resulting posterior probability estimates for the hidden state variable, i.e. 
$$ \hat P(S_t=1 \vert \tilde{\textbf{u}}_T) := \sum_{r=1}^{R} {S}_t^{r,MCMC}, $$
for $R$ independent MCMC samples, are plotted as dotted points in Figure \ref{ex_rvine_switch}. These Bayesian estimates follow those obtained from the EM algorithm closely, showing only a little bit more variability. 

\subsubsection{Identifying crisis regimes}\label{sec_crisis}
\begin{figure}[H]
\centering
\includegraphics[width=0.9\textwidth]{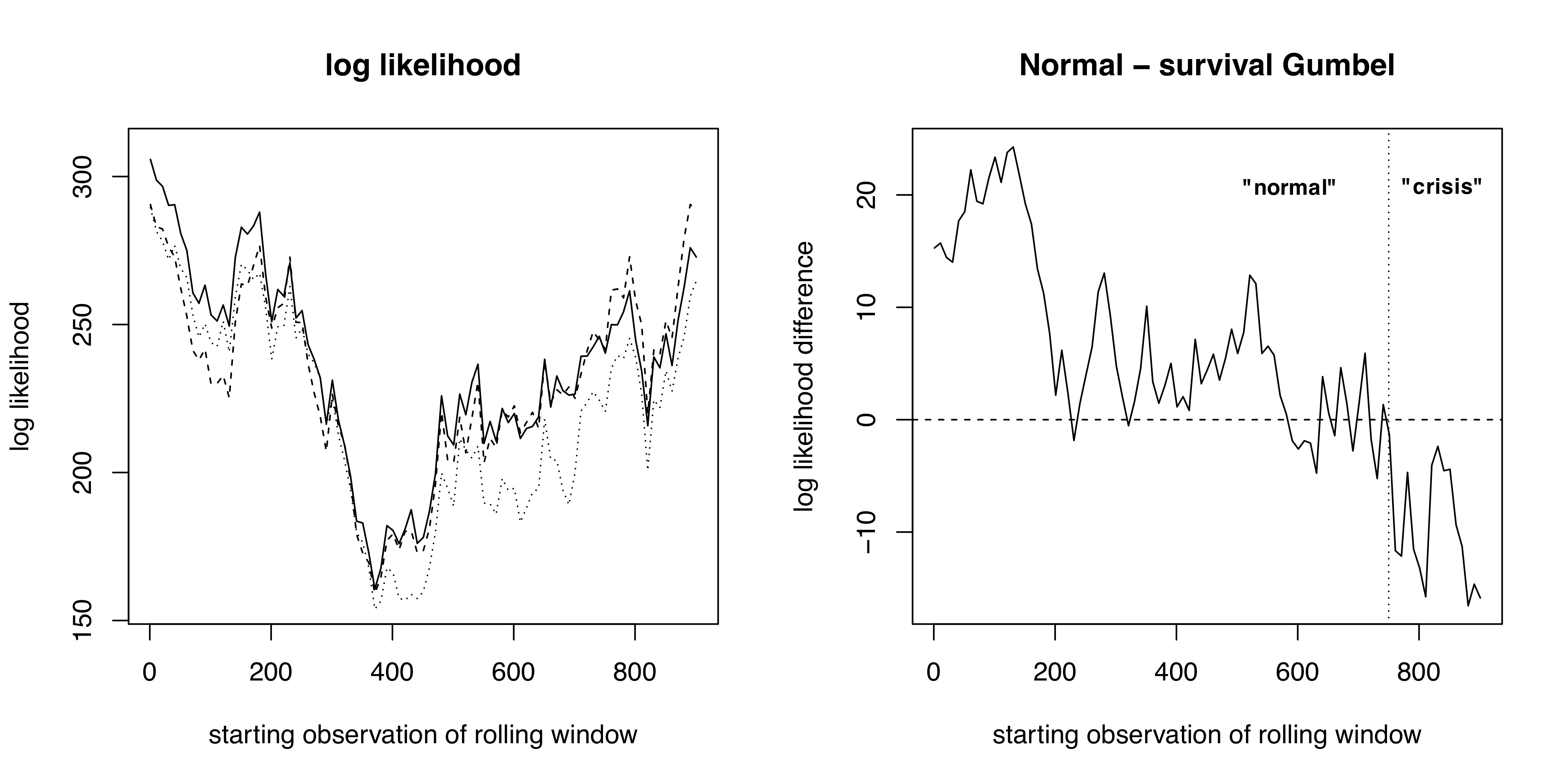}
\vspace{-.3cm}
\caption{Left-panel: log likelihood values resulting from fitting R-vines with normal (solid), survival Gumbel (dashed) and Gumbel (dotted) copulas. Right panel: difference between the values for the normal and Gumbel model, we indicate periods to which the structures for Model (3) are fitted.}
\label{rolling_window_ll}
\end{figure}

Having identified parameter switches in an R-vine copula model for our data set, we will now try to identify switches in the overall dependence structure.
Since there is empirical evidence that dependence structures can change in times of crisis (c.f. \citeN{longin1995}, \citeN{ang2002} or \citeN{garcia2011}) and since tail dependencies become more important in times of extremal returns, we want to select two different R-vine structures using the aforementioned procedures. 
To do so, we start with a rolling window analysis, selecting and fitting R-vine models to a rolling window of 100 data points. To reduce model complexity, we decide to work again with truncated and simplified R-vines, resulting in a sufficiently flexible and parsimonious model. The copulas on the first tree were chosen to be either all Gaussian, Gumbel or survival Gumbel. The copulas on the second tree were set to Gaussian and the R-vines were truncated after this second tree. The resulting rolling log likelihoods from conducting this analysis are given in Figure \ref{rolling_window_ll}. Note that for AIC (BIC) comparison this is sufficient since the number of parameters remains the same in all models considered.

We see that, while the range of overall likelihood estimates is similar, the Gaussian model tends to give the best fit, i.e. highest log likelihood, (left panel of Figure \ref{rolling_window_ll}) over the whole data set. However, the survival Gumbel model starts to outperform the Gaussian model towards the end of the observation period (right panel of Figure \ref{rolling_window_ll}). 
Furthermore, the survival Gumbel model, in which the exchange rates taken into consideration are assumed to be lower tail dependent, tends to outperform the model with standard Gumbel copulas, corresponding to upper tail dependence.
This is in accordance with the observation that the financial crisis during the observation period originated in the dollar area, quickly spreading to the world economy but with different severity e.g. to the other developed countries and the developing countries. Because of this, cash flows out of the dollar area, resulting in higher FC/US exchange rates, tend to be less extremely correlated than cash flows into the dollar area to settle liabilities denominating in US dollar, which results in more lower than upper tail dependence.
Given these observations from the rolling window analysis, we decide to select R-vine copulas as follows: For Regime 1, the tree structure ($\mathcal{V}_1$) is again fitted to the whole data set, but we use only Gauss copulas as bivariate building blocks. Since parameter estimates on the higher trees indicate week dependence or even conditional independencies, we truncate the R-vine copula after the second tree.
To determine a second structure ($\mathcal{V}_2$) we apply our outlined techniques to the time frame from July 10, 2008 to December 3, 2008, where the likelihood of the survival Gumbel model starts to be higher than the likelihood of the Gaussian model. Doing so, we also capture many high-impact events of the financial crisis. 
For the copulas associated to the R-vine structure we consider
\storestyleof{itemize}
\begin{listliketab}
  \begin{tabular}{p{.16\textwidth} p{.83\textwidth}}
  Model (2a) & survival Gumbel copulas on the first tree to capture strong dependencies for negative returns. \\
  Model (2b) & Gumbel copulas on the first tree to capture dependencies in the upper tail. \\
  Model (2c)  &Student-t copulas on the first tree to cover symmetric tail dependencies. \\
  \end{tabular}
  \vspace{-.8cm}
\end{listliketab}

The copulas corresponding to edges on the second tree are again chosen to be Gaussian and we truncate after Tree 2.
While the survival Gumbel model is preferred in the rolling window analysis, we include Models (2b) and (2c) to investigate the impact of different tail dependencies. Note that Model (2c) where we use Student-t copulas in the second structure is close to considering an R-vine model where the structure is kept fixed and only the parameters are subject to regime switches if the degrees of freedom parameters are high. In this case, we are left with two different (truncated) R-vine structures where all copulas are Gaussian.
Since an R-vine where all bivariate building blocks are Gaussian leads to a multivariate Gaussian copula, the possible dependencies in the two regimes do only differ because of the truncation ofter the second tree.
Given the fact that limiting the range of copula families which can be associated to the R-vine as in Models (2a) - (2c) necessarily leads to lower AIC values we do further include a Model (3) which also has different dependence structures in both regimes but where all bivariate copulas are selected using the stepwise AIC criteria.
To be more precise, the R-vine structure $\mathcal{V}_3$ together with the corresponding copulas for the "crisis" regime is selected by applying the stepwise selection procedure to the part of our dataset where the rotated Gumbel copula is outperforming the normal copula in the scenario analysis, which corresponds approximately to the last 250 observations (annotated with "crisis" in the right panel of Figure \ref{rolling_window_ll}). The R-vine structure for the "normal" regime, again with corresponding copulas is identified from the remaining data points, it is the same as for the "normal" regime in Models (2a) - (2c), $\mathcal{V}_1$.
While Models (2a) - (2c) are designed with specific dependence structures in order to investigate the changes of dependence present in the data more closely, Model (3) makes use of the full modeling flexibility of MS R-vines in order to provide a close fit to the data at hand.
We employ the stepwise EM-procedure to fit MS models with the selected regimes to our data set.
The resulting smoothed probabilities inferred from the hidden state variable for being in the non-Gaussian regime using Models (2a) - (2c) are given in Figure \ref{ex_crisis}.

While the overall strength of dependence modeled in the two regimes (judging by the fitted values of Kendall's $\tau$, see Tables \ref{param_model2_1} and \ref{param_model2_2} of Appendix \ref{appendix_rvine_structure}) is similar for all three models, the results for Model (2c) with Student-t copulas are close to the results of Model (1), whereas the other two differ significantly.
This was expected, since the model with t-copulas is close to a Gaussian copula model with regime switching parameters.
Analyzing the estimated Kendall's $\tau$ further, we find that in Model (2c) the Kendall's $\tau$ between the Japan-US and the India-US exchange rate indicates negative dependence (see Table \ref{param_model2_1}). Since Gumbel and survival Gumbel copula models exhibit only positive dependence, this cannot be captured in Model (2a) or (2b), respectively.
Replacing the copula for this bivariate margin by a Gauss copula (we refer to the resulting modification of model (2a) as (2$\text{a}^\star$)) so that it captures the negative dependence does however not significantly change the posterior estimates for the hidden state variable. 
This means that the observed difference in the behavior of Models (2a) and (2b) as compared to Model (2c) cannot be explained by the lack of Gumbel and Gumbel survival copulas to allow for negative dependence.
Instead these models tend to be preferred during specific times of high impact events of the financial crisis, where the bivariate dependence structures are closer to the dependence structure of a Gumbel copula, as indicated in Figure \ref{ex_crisis} where some important events are annotated.

\begin{figure}[!ht]
\centering
\includegraphics[width=0.9\textwidth]{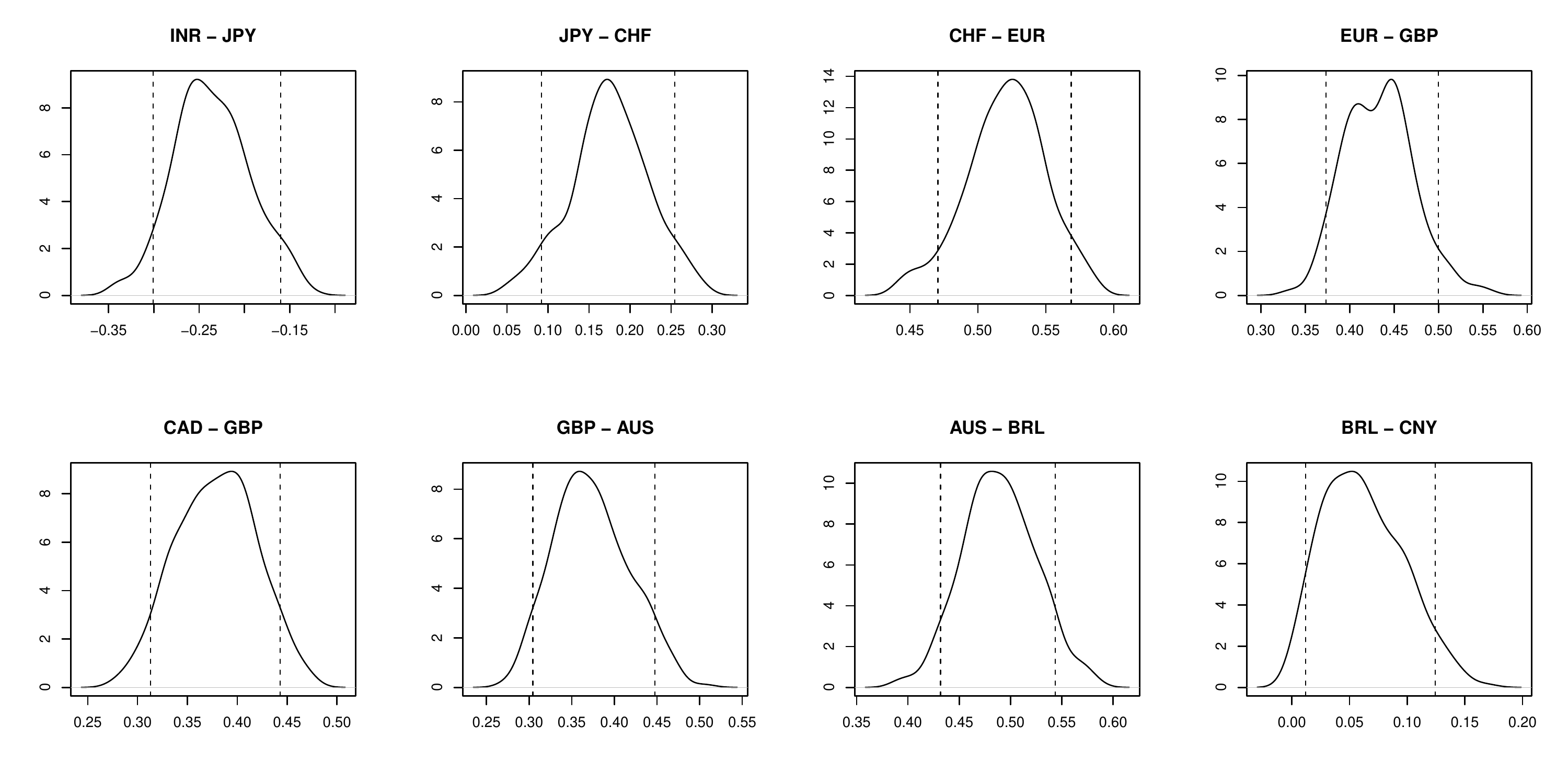}\caption{Estimated marginal posterior densities in the crisis regime of Model (2$\text{a}^\star$) with 90\% CIs. The plotted densities correspond to the (unconditional) copulas associated to Tree 1 of the vine $\mathcal{V}_2$.  }
\label{posterior_histograms}
\end{figure}

While obtaining a sample from the posterior distribution of the latent state variable is a byproduct from the Bayesian estimation procedure, its main purpose is to characterize the joint posterior distribution of the model parameters in the R-vine copula. 
Figure \ref{posterior_histograms} shows histogram plots of several marginal posterior densities for the copula parameters in the crisis regime in Model (2$\text{a}^\star$). As we can see, the parameter value of $\tau_{INR-JPY}=0$ which would correspond to independence is nowhere near a $90$ or $95$ percent CI, the dependence is significantly negative. For the copula between Brazil-US and China-US in contrast, the parameter values in our posterior sample are all close to $0$, which means that the two time series are only weekly dependent or maybe even independent. 

\begin{minipage}[c]{\textwidth}
\centering
\includegraphics[width=0.9\textwidth]{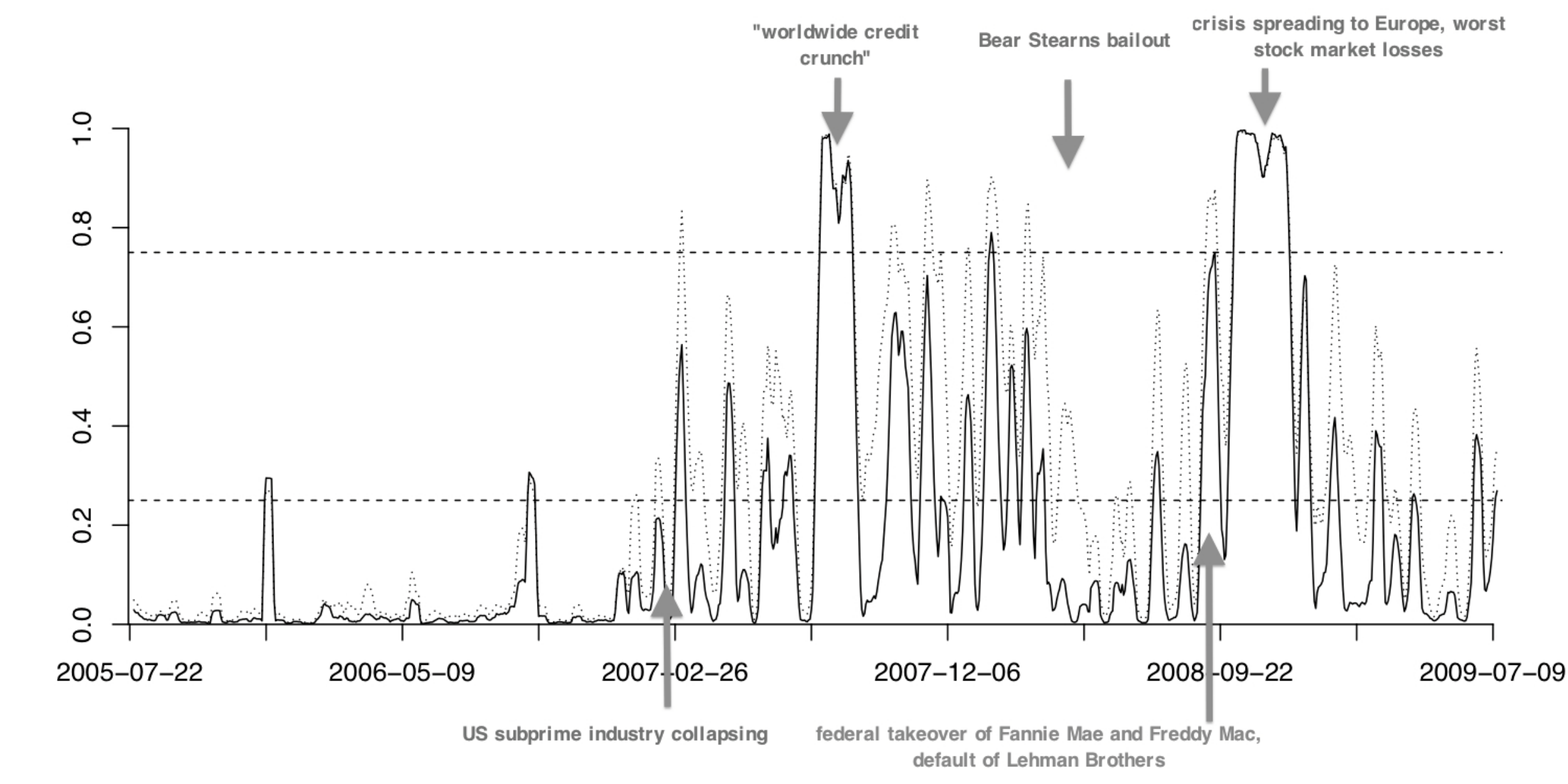}

\vspace*{-1.0cm}
\includegraphics[width=0.9\textwidth]{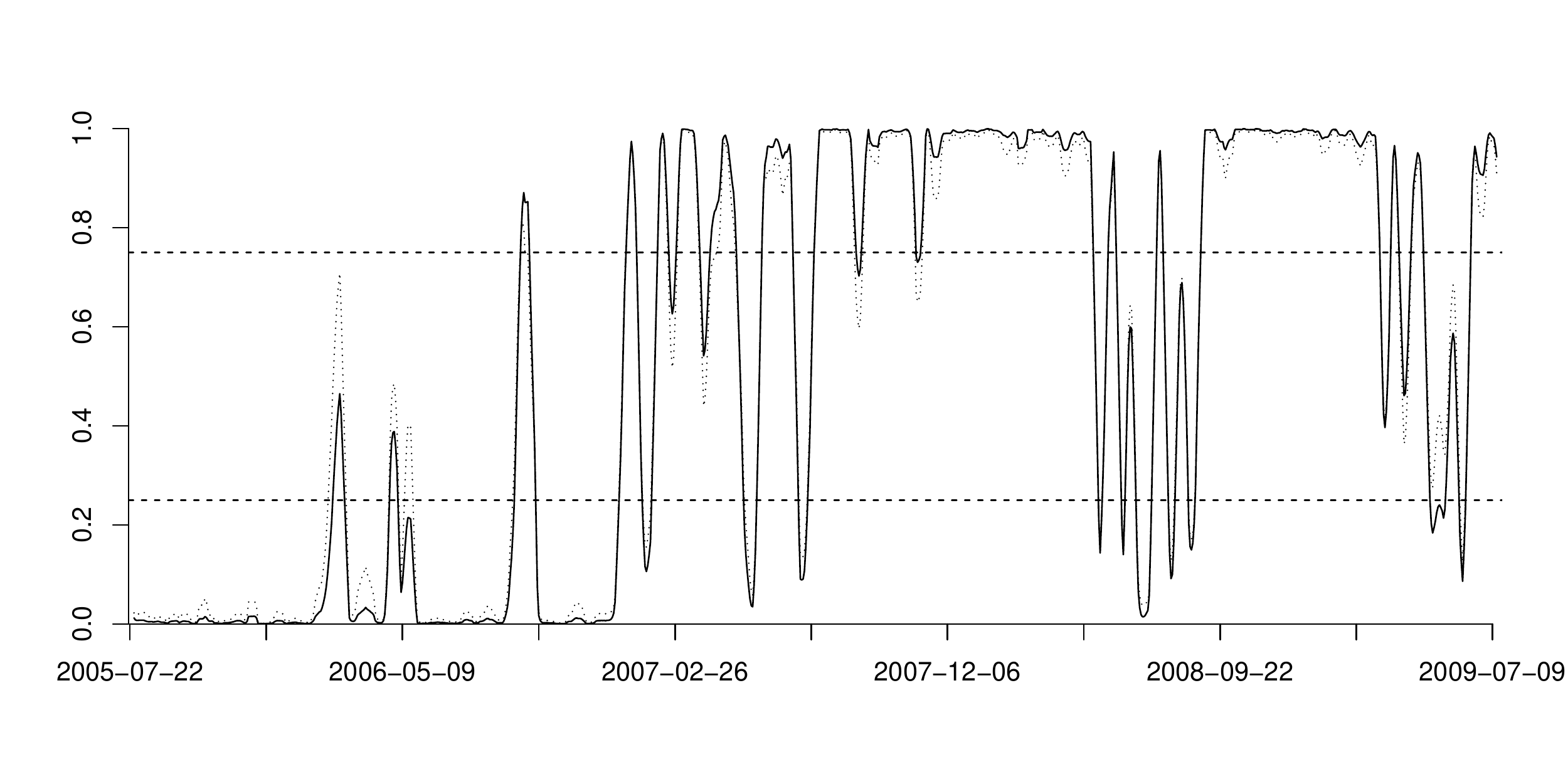}

\vspace*{-1.0cm}
\includegraphics[width=0.9\textwidth]{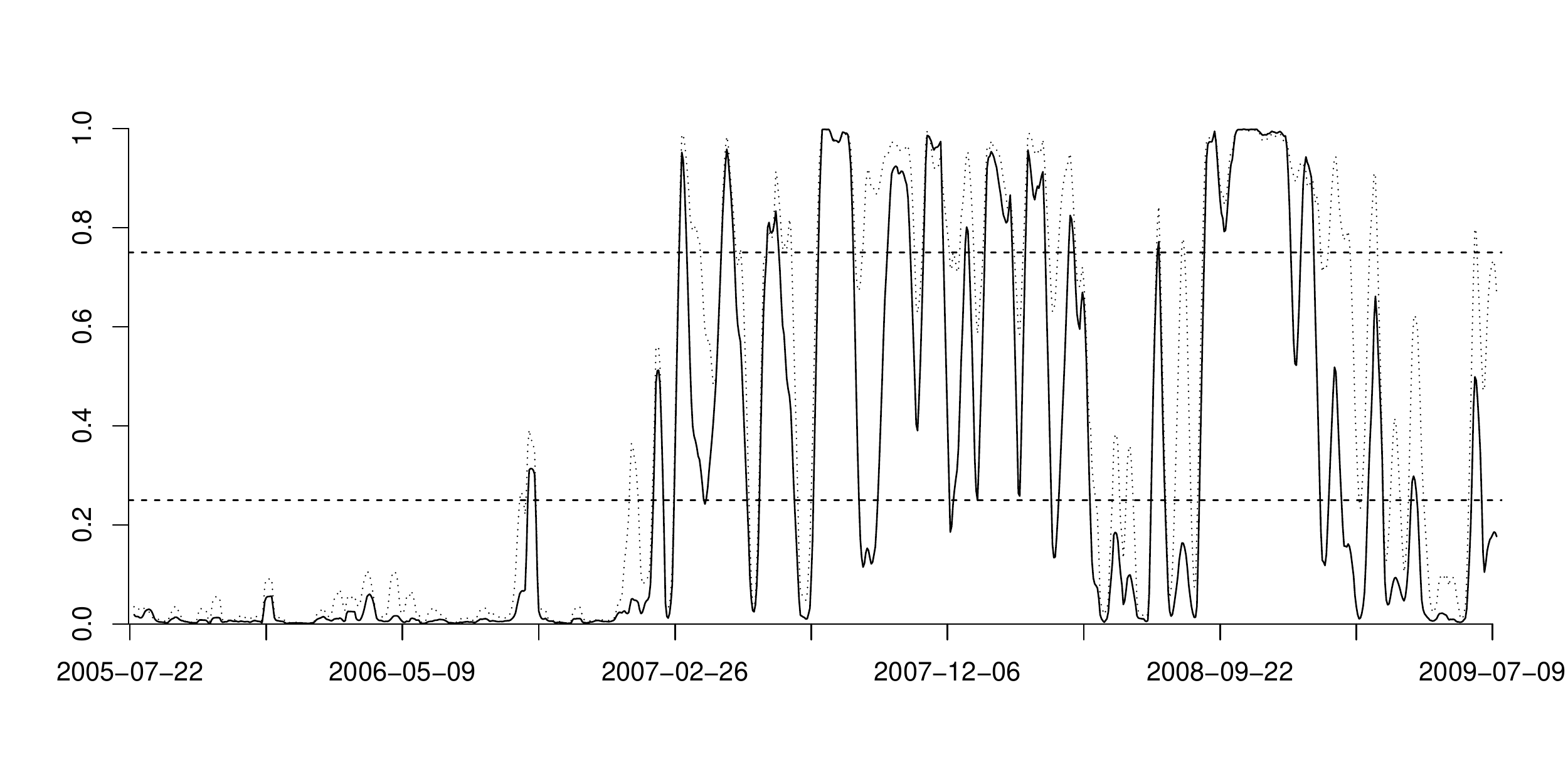}

\vspace*{-.5cm}
\captionof{figure}{Smoothed probabilities that the hidden state variable indicates the non-Gaussian regime. From top to bottom: model (2a) with Gumbel survival copulas, model (2b) with Gumbel copulas and model (2c) with Student-t copulas. The solid lines correspond to EM estimates while Bayesian MCMC estimates are dotted, high impact events of the financial crisis are annotated in the upmost graph.}
\label{ex_crisis}
\end{minipage}

\subsubsection{Model comparison}
Having discussed the stylized features of the different R-vine models we have taken into consideration for the exchange rate data, we want to compare them in terms of their fit to the observations at hand.
For this, we rely on in-sample methods, and use our Bayesian Gibbs sampling procedure to calculate the deviance information criterion (DIC) which has been proposed by \shortciteN{spiegelhalter2002}. 
\begin{table}[]
\begin{center}
\begin{tabular}{ c c c  c  c  c  c  c  }
Model & (1) & (2a) &($\text{2a}^\star$) & (2b) & (2c) & (3) & no MS \\
\hline
DIC & -4398 & -4280 & -4312 & -4199 & -4346 & -4430 & -4146 \\
\end{tabular}
\caption{DIC values for the different (regime switching) R-vine models that have been considered. Lower values indicate a better fit of the model to the data.}
\label{dic_table}
\end{center}
\end{table}
Table \ref{dic_table} shows DIC values for all models under investigation, calculated using the Gibbs sampling procedure. For comparison purposes, we also include an R-vine model without MS, but where the vine tree structure has not been truncated after tree 2. The first two trees of this structure correspond to Structure $\mathcal{V}_1$.

Although the full R-vine model has 36 parameters and the MS R-vine models where we use truncated vines and one parametric pair copulas only have 32, even the worst MS R-vine copula outperforms the model without Markov structure, which clearly supports the use of models with time varying dependence in this context.
The DIC values further show that in terms of in-sample fit, the model with standard Gumbel copulas in the crisis regimeis outperformed by the other models, which was to be expected from the rolling window analysis.
Since the copulas in Model (1) were chosen maximizing pairwise AIC, it outperforms the models where we restricted the choice of copulas.
The best-performing model however is Model (3), where the copula families were chosen using pair-wise AIC but the R-vine structure differs between the regimes. This shows that MS models for all components of the dependence structure are more suitable for this kind of data than models where only the copula parameters are varying over time.

\subsection{Eurozone country indices} \label{sec_applications_index}
\begin{figure}[!ht]
\centering
\includegraphics[width=\textwidth]{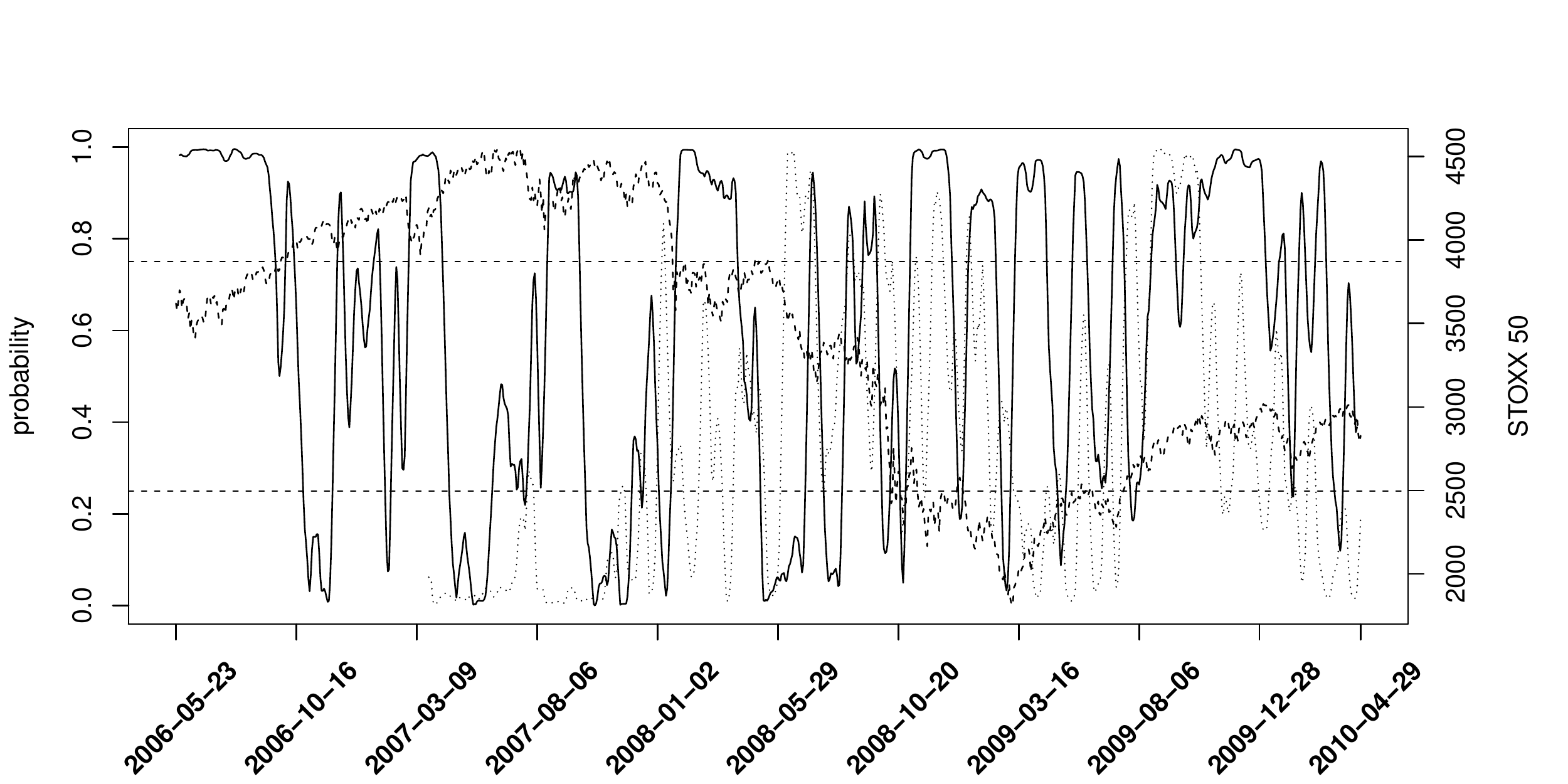}\caption{Smoothed probabilities (EM algorithm) that the latent variable indicates the presence of the high dependence regime in the STOXX data set (solid) vs. smoothed probabilities that the latent state variable indicates the presence of the crisis regime in the exchange rate data set (dotted) vs. quoted STOXX 50 values (dashed, right y-axis).}
\label{STOXX_figure}
\end{figure}

To illustrate to the interested reader that MS can be detected in the dependence structure of financial data sets of various kind, we briefly outline two more applications. 
Our next data set consists of daily log-returns for 5 stock indices of the eurozone: the German DAX, the French CAC 40, the Dutch AEX, the Spanish IBEX 35 and the Italian FTSE MIB. We consider 985 observations between May 22, 2006 and April 29, 2010. The data has already been analyzed by \citeN{brechmann2011}, we refer to their Appendix A for the determination of marginal time series models. Since we only want to investigate the presence of MS, we consider the simplest possible model, having the same tree structure and copula families in both regimes. 
With the same technique as in Section \ref{ex_Rvine} (Model (1)) we fit this R-vine with only MS-parameters to the data (see Figure \ref{stoxx_tree_figure}, Appendix \ref{appendix_rvine_structure}). Here, the two regimes differ significantly in terms of the dependency strength they describe (see Table \ref{stoxx_tree}, Appendix \ref{appendix_rvine_structure}). While the copula parameters in the first regime correspond to values of Kendall's $\tau$ of $0.6-0.75$, the second regime has values between $0.75$ and $0.85$.
For comparison, the probabilities for the high dependence regime are plotted in Figure \ref{STOXX_figure} together with the probabilities for the "crisis" regime in Model (2a) for the exchange rate data and the quoted values of the STOXX 50. It shows, that the high dependence regime for eurozone country indices becomes more relevant during times of a weakening economy and that it's presence is positively correlated with the presence of the "crisis" regime in Model (2a) of the exchange rate data, reflecting the interrelations between foreign exchange markets and stock markets.

\subsection{Selected German Stocks}\label{sec_applications_dax}

\begin{figure}[!ht]
\centering
\includegraphics[width=\textwidth]{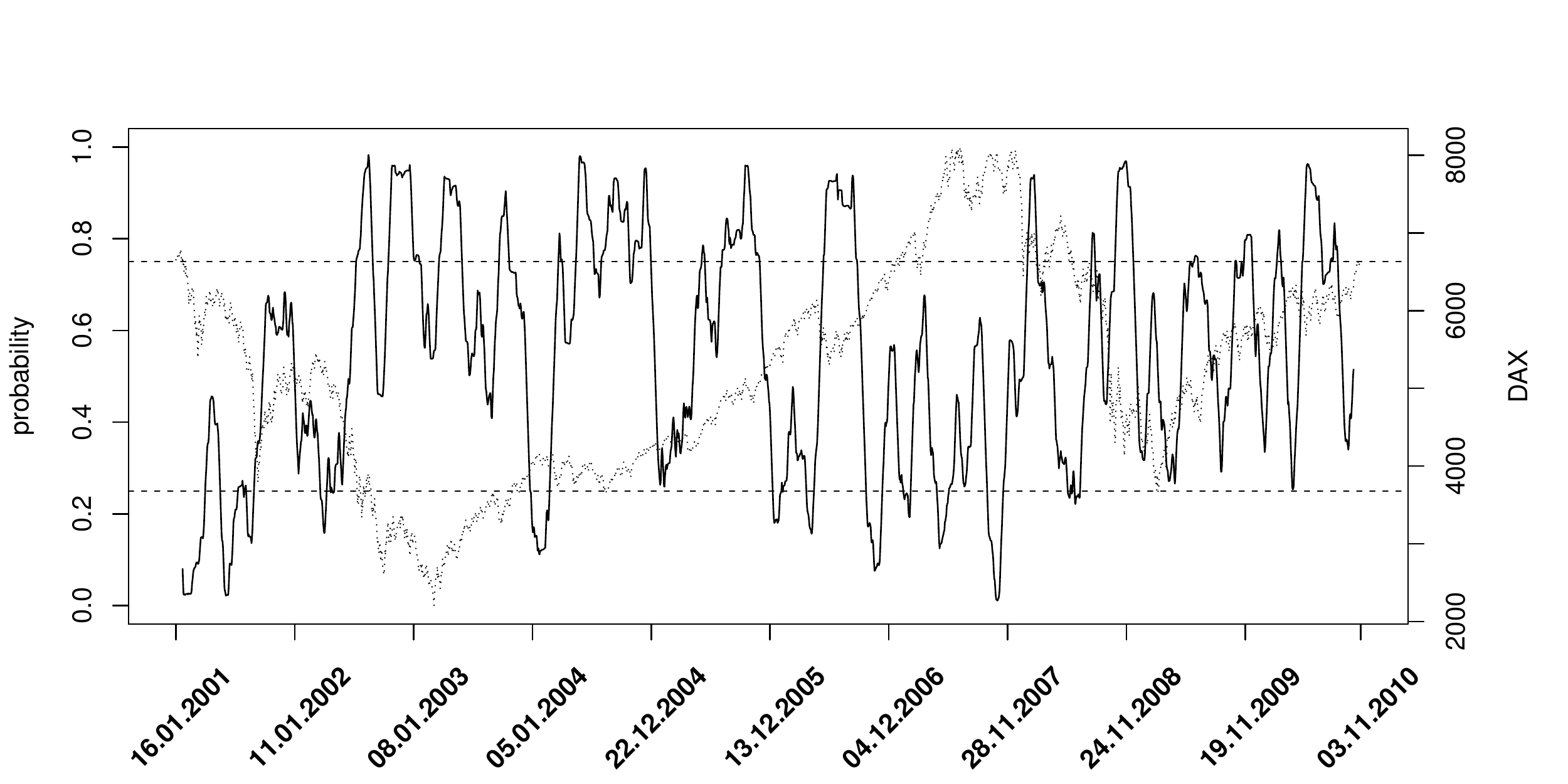}\caption{Smoothed probabilities (EM algorithm) that the high dependence regime is present in the German stock data (solid, left y-axis) vs. quoted values of the DAX (dotted, right y-axis).}
\label{DAX_figure}
\end{figure}

As a third application, we consider daily log-returns of the 10 stocks in the German stock index DAX with the highest market capitalization on October 25, 2010, namely: Allianz (ALV), BASF (BAS), Bayer (BAY), Daimler (DAI), Deutsche Bank (DBK), Deutsche Telekom (DTE), E.ON (EOA), RWE, SAP, Siemens (SIE).
The observed time period ranges from January 17, 2001 to November 4, 2010, resulting in 2494 daily observations. The analysis of the one dimensional margins and the transformation to copula data has been performed in \citeN{stoeber2011}. For demonstration purposes, we again choose the simplest possible model.
Employing again the technique of Sections \ref{ex_Rvine} and \ref{sec_applications_index}, we fit an R-vine with only MS-parameters (Figure \ref{dax_trees_app}, Appendix \ref{appendix_rvine_structure}). As for the indices, we identify one regime with weaker (Kendalls $\tau$ on the first tree $\sim 0.4$) and one regime with stronger ($\tau \sim 0.6$) dependencies (see Table \ref{dax_app_table}, Appendix \ref{appendix_rvine_structure}. 
The probability over time that this second regime is present is plotted in Figure \ref{DAX_figure}.
It shows that the probability for the high dependence regime and the absolute index values are clearly negatively correlated for the period from 2001 to 2008. Since the beginning of 2009 we observe that, despite rising index values, also the probabilities are in an upward trend. 
The negative correlation at the beginning suggests an interpretation of the high dependence regime as a kind of "crisis" regime, which governs the dependence during times of a weak economy. In this perception, the upward trend in the state probability since 2009 is an indicator for the remaining of the German economy in a "crisis" state and a growing destabilization due to the public debt problems in the eurozone.

\section{Discussion and Outlook}\label{sec_conclusion}

This paper provides a detailed investigation of estimation methods for Markov switching regular vine models, which constitutes a significant contribution towards making time-varying copula models a standard tool for the description of multivariate time series. 
Allowing flexible pair copula constructions, which can account for the stylized facts observed in the dependence among financial time series, to vary over time, we create a model which is limited mainly by its need for efficient computational treatment.
The quick EM algorithm, based on the step-wise estimator for R-vine copulas, which we have introduced, allows to perform ML inference in almost arbitrary dimensions. For a more thorough study of the resulting model parameter estimates and their uncertainty, we further introduced a Bayesian inference procedure which has been shown to correctly capture the true model in simulated data.

Our findings in empirical applications illustrate that regime switches are present in many financial data sets, including exchange rates, index returns and stock returns. In particular, regime switching models constitute possible tools for the accurate description of changes in dependence during times of crisis. In this context we have also demonstrated a possible model selection heuristic for the context of regime switching dependence models using the R-vine selection procedures pioneered by \shortciteN{dissmann2011}. 

While we believe that the methods for parameter estimation presented here will satisfy most statisticians and practitioners needs, improvements in model selection are still desirable. It is computationally intensive to perform model selection by computing the DIC from a sufficiently large sample of the parameters' posterior distribution, therefore model selection will continue to be a topic of ongoing research. In particular, we will investigate how the rolling window methods can be improved and automatized using change point detection.

\textbf{Acknowledgement} The numerical computations were performed on a Linux cluster
supported by DFG grant INST 95/919-1 FUGG. The first author further acknowledges financial support by a research stipend from Allianz Deutschland AG. Both authors want to thank the participants of the 4th Workshop on Vine copulas, Munich, 2011, IWSM, Valencia, 2011 and ISI, Dublin, 2011, where preliminary results of this paper have been presented, for their helpful comments.
\appendix
\appendixpage
\addappheadtotoc

\section{Step-wise estimation of R-vine copula parameters}\label{appendix_stepwise}

In this appendix, we will shortly describe the step wise parameter estimation method for an R-vine copula for the case where observations are subject to different weights. Here, the function to be maximized is a weighted log likelihood
\begin{equation*}
l_{\tilde{\bomega}_T}(\btheta):=\sum_{t=1}^T log(c(\textbf{u}_t \vert \mathcal{V}, \textbf{B}, \btheta )) \cdot \omega_t,
\end{equation*}
where $(\textbf{u}_t)_{t=1,\dots,T}$ is a series of independent observations and $\{ \omega_t, t=1,\dots,T\}$ are the corresponding weights.

For the case where weights are constant, this estimation procedure has already been applied by several authors among which we want to particularly note \citeN{haff2010} and \shortciteN{CzadoSchepsmeierMin2011}. For more details, in particular on asymptotic properties and further references we refer to the paper of \citeN{haff2010}.
In the context of MS models however, the weights to be attributed to individual observations are determined in the Expectation Step of the EM algorithm and are non-constant therefore, in particular $\omega_t=\left(\bOmega_{t\vert T}(\btheta^l)\right)_{s_t}$ at iteration $l$.

Before we can come to the algorithm itself, we need to introduce some more notation. Evaluating the density of an R-vine copula using Expression (\ref{rvine_density}) involves arguments \begin{gather*}
F(x\vert {\textbf{x}})=\frac{\partial C_{x,{x}_j \vert {\textbf{x}}_{-j}}\left(F(x\vert {\textbf{x}}_{-j}),F({x}_j\vert {\textbf{x}}_{-j})\right)}{\partial F({x}_j\vert {\textbf{x}}_{-j})},
\end{gather*}
where ${\textbf{x}}_{-j}$ is short notation for the vector where the $j$-th component of ${\textbf{x}}$ (${x}_j$) has been removed.
Following Aas et al. (2009), we will denote this conditional (copula) distribution function in the bivariate case when $X_1=U_1$ and $X_2=U_2$ are uniform, i.e. jointly distributed according to some copula $C$, by $h(u;v,\boeta,t)$. Then,
\begin{gather*}
h(u_1;u_2,\boeta,t)=F(u_1\vert u_2)=\frac{\partial C_{u_1,u_2}(u_1,u_2;\boeta,t)}{\partial u_2},
\end{gather*}
where the second argument $u_2$ corresponds to the conditioning variable and $(\boeta,t)$ denotes the set of parameters $\boeta$ together with the copula type $t$ for the copula $C_{u_1,u_2}$ of the joint distribution. The corresponding copula density will be denoted by $c(u_1,u_2;\boeta,t)$

Using this, the estimation algorithm is given by the steps outlined in Algorithm \ref{stepwise_estimates}. We use the convenient matrix notation for this computational purpose and our algorithm determines the correct copula for each step.
As in Section \ref{sec_model_rvine}, we denote the parameters and copula types corresponding to the copula associated to edge $e$ with $\boeta_{j(e),k(e)\vert D(e)}$ and $t_{j(e),k(e)\vert D(e)}$ respectively.

\begin{algorithm}[H]
\caption{Obtain stepwise parameter estimates for an R-vine copula.}
 \label{stepwise_estimates}
\begin{algorithmic}
\REQUIRE R-vine copula specification in the form of an R-vine matrix with corresponding sets of copula parameters / types, and observations $(\textbf{u}_1,\dots,\textbf{u}_T)$, $\textbf{u}_t=(u_{1t},\dots,u_{dt})$ from the R-Vine copula distribution with corresponding weights $\{ \omega_t, t=1,\dots,T \}$
\STATE Define $\mathbbm{m}_{ki}=max\{m_{k,i},...,m_{n,i}\}$
\STATE Set  $v^{direct,t}_{d,i} := u_{it}  \ \ \ \ \   t=1,\dots,T \ \ i=1,\dots,d$
\FOR {i in d-1,...,1}
\FOR {k in d,...,i+1}
\STATE $zr_1^t := v^{direct,t}_{k,i} \ \ \ \ \ t=1,\dots,T$
\IF{$\mathbbm{m} = m_{k,i}$}
\STATE $zr_2^t := v^{direct,t}_{k,(d-m+1)} \ \ \ \ \  t=1,\dots,T$
\ELSE
\STATE $zr_2^t := v^{indirect,t}_{k,(d-m+1)}  \ \ \ \ \  t=1,\dots,T$
\ENDIF
\STATE Obtain $p_{k,i}$ via maximization of\\ 
\qquad \qquad $\sum_{t=1}^T\left[ log(c(zr_1^t, zr_2^t; \boeta_{m_{k,i},m_{i,i}\vert m_{k+1,i}, m_{d,i}},t_{m_{k,i},m_{i,i}\vert m_{k+1,i}, m_{d,i}})) \cdot \omega_t\right]$				
\STATE Set $v^{direct,t}_{k-1,i} := h(zr_1^t, zr_2^t ;\boeta_{m_{k,i},m_{i,i}\vert m_{k+1,i}, m_{d,i}},t_{m_{k,i},m_{i,i}\vert m_{k+1,i}, m_{d,i}} ) \ \ \ \ \   t=1,\dots,T$
\STATE Set $v^{indirect,t}_{k-1,i} := h(zr_2^t, zr_1^t; \boeta_{m_{k,i},m_{i,i}\vert m_{k+1,i}, m_{d,i}},t_{m_{k,i},m_{i,i}\vert m_{k+1,i}, m_{d,i}} ) \ \ \ \ \   t=1,\dots,T$
\ENDFOR
\ENDFOR
\end{algorithmic}
\end{algorithm}

\section{Relative bias and MSE for parameter estimates in the simulation study}\label{appendix_simstudy}
The following tables show the relative bias and relative MSE for the parameters of two selected scenarios in the simulation study (Scenario 2 \& Scenario 4). For comparison purposes all copula parameters have been transformed to the Kendall's $\tau$ level.

Notice the large bias for the posterior mean estimate of the second Markov chain parameter in Scenario 4 (Table \ref{simstudy_mcbias}) where identification issues where observed. In this case the Gibbs sampler with objective priors fails to capture the underlying Markov structure correctly and the Bayesian procedure needs to be started with strong subjective prior beliefs.

In general, we observe that the estimation error in the second and third tree is higher than on the first tree and that the uncertainty in the Gumbel regime, from which less realizations are included in the data set, is higher than in the Gaussian regime.

\begin{minipage}{\textwidth}
\includegraphics[width=0.9\textwidth]{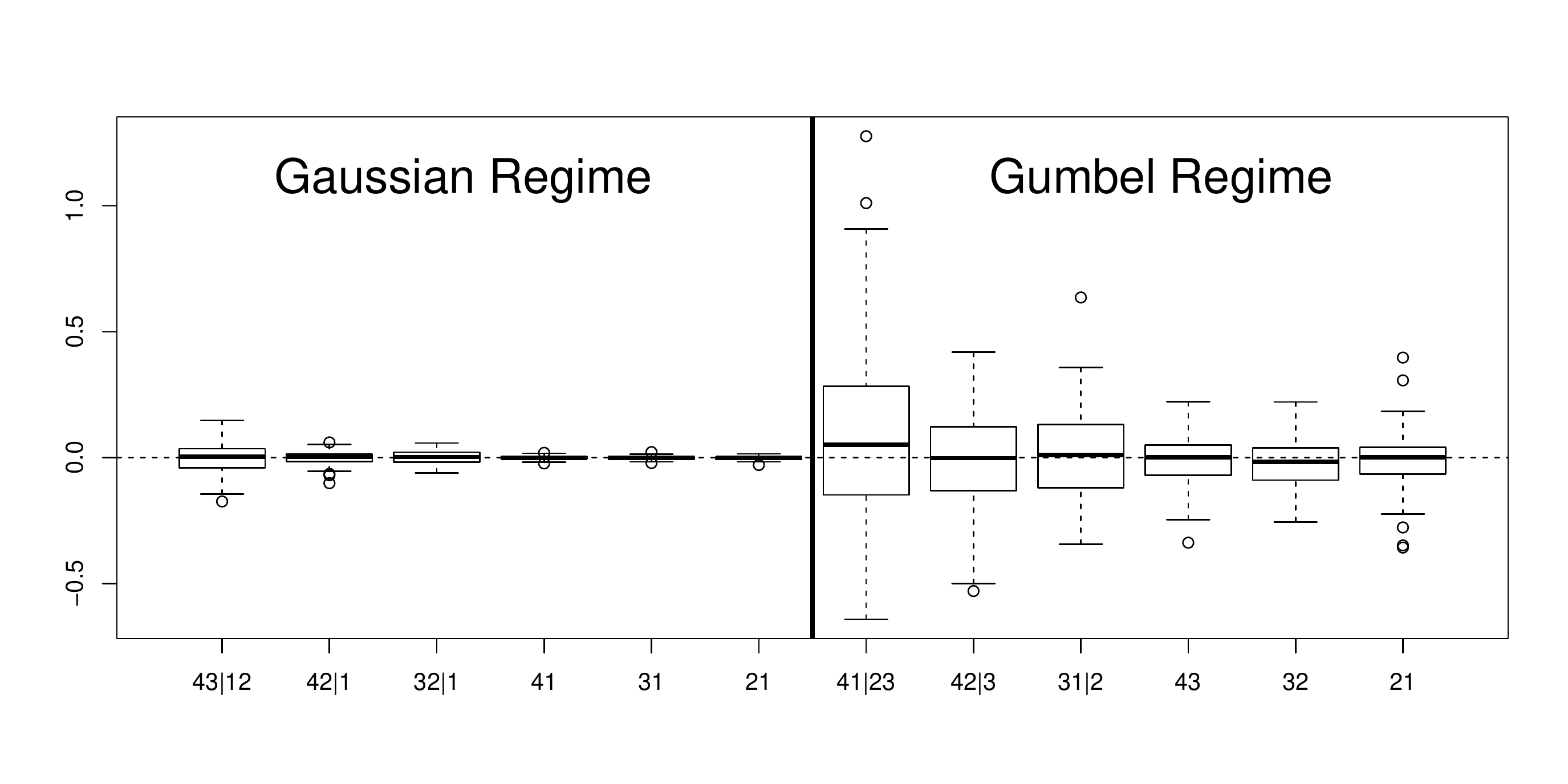}
\centering
\begin{tabular}{c c  c  c  c  c  c }
Gaussian Regime & $\tau_{43 \vert 12}$ & $\tau_{42 \vert 1}$ & $\tau_{32 \vert 1}$ & $\tau_{41}$ & $\tau_{31}$ &$\tau_{21}$ \\
\hline
relative bias  & -2.1 $\cdot 10^{-3}$ & -1.7 $\cdot 10^{-3}$ & 7.2 $\cdot 10^{-4}$ & -8.1 $\cdot 10^{-4}$ & -1.1 $\cdot 10^{-3}$ & -1.9 $\cdot 10^{-3}$ \\
relative MSE   & 1.4 $\cdot 10^{-3} $ & 4.3 $ \cdot 10^{-4}$ & 4.3 $\cdot 10^{-4}$ & 4.6 $\cdot 10^{-5}$ & 4.1 $ \cdot 10^{-5} $ & 5.5 $\cdot 10^{-5} $ \\
\hline
\hline
Gumbel Regime & $\tau_{41 \vert 23}$ & $\tau_{42 \vert 3}$ & $\tau_{31 \vert 2}$ & $\tau_{43}$ & $\tau_{32}$ &$\tau_{21}$ \\
\hline
relative bias  & 
9.5 $\cdot 10^{-2}$ & 5.2 $\cdot 10^{-3}$ & 1.7 $\cdot 10^{-2}$ & -1.1 $\cdot 10^{-2}$ & -1.6 $\cdot 10^{-2}$ & -1.0 $\cdot 10^{-2}$ \\
relative MSE   & 1.3 $\cdot 10^{-2} $ & 7.2 $ \cdot 10^{-3}$ & 5.6 $\cdot 10^{-3}$ & 3.3 $\cdot 10^{-3}$ & 3.2 $ \cdot 10^{-3} $ & 3.6 $\cdot 10^{-3} $ \\
\end{tabular}
\captionof{table}{Scenario 2: Relative error of Kendall's $\tau$ estimates (top figure) and relative bias / MSE for the Gaussian regime (upper table) and the Gumbel regime (lower table), respectively.}
\end{minipage}

\begin{minipage}{\textwidth}
\begin{minipage}{0.5\textwidth}
\centering
\includegraphics[width=0.8\textwidth]{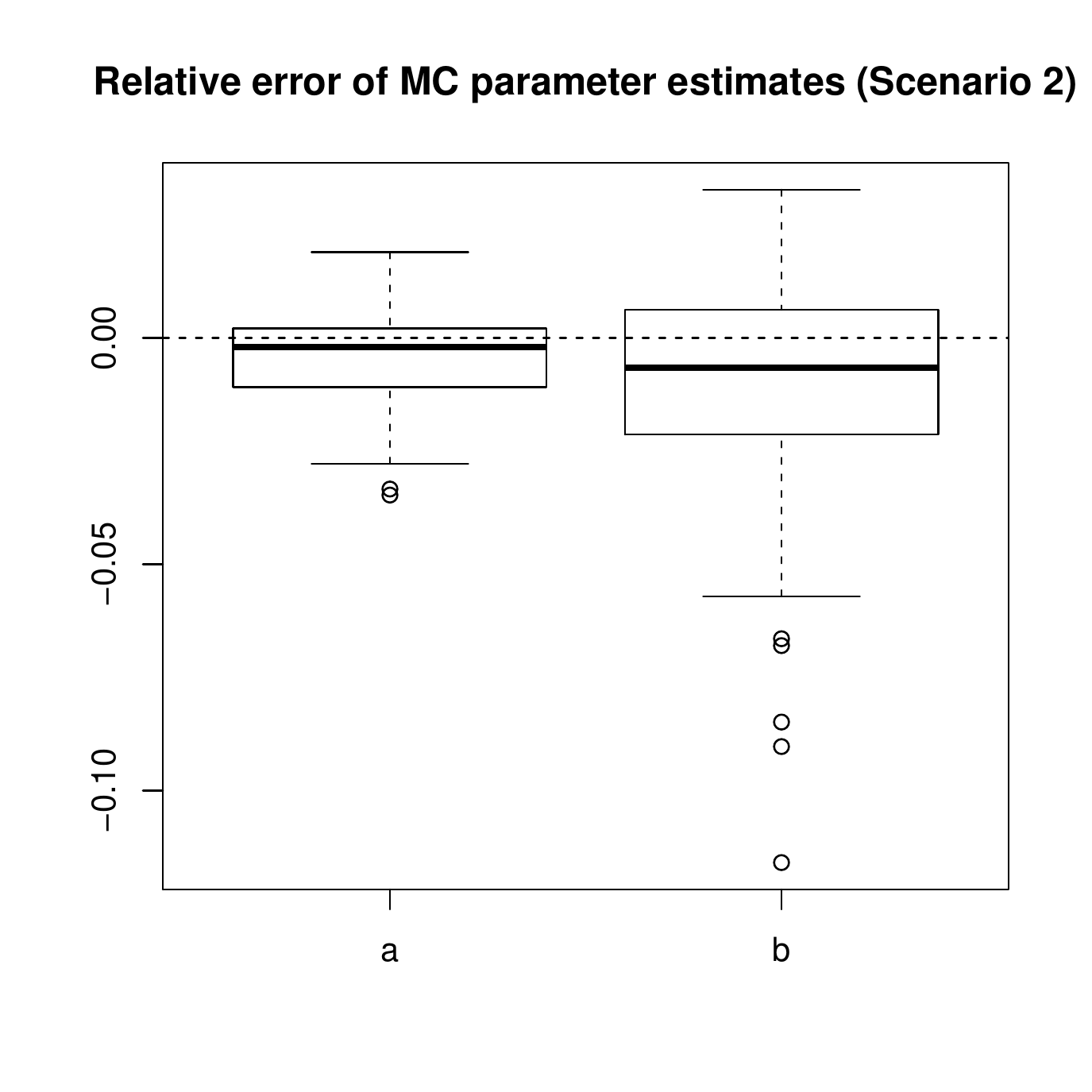}\vspace{-.8cm}
\begin{tabular}{c c  c}
MC parameters &$a$ & $b$ \\
\hline
relative bias  & -3.7 $\cdot 10^{-3}$ & -1.0 $\cdot 10^{-2}$  \\
\hline
relative MSE   & 1.4 $\cdot 10^{-4} $ & 6.5 $ \cdot 10^{-4}$  \\
\end{tabular}
\end{minipage}
\begin{minipage}{0.5\textwidth}
\centering
\includegraphics[width=0.8\textwidth]{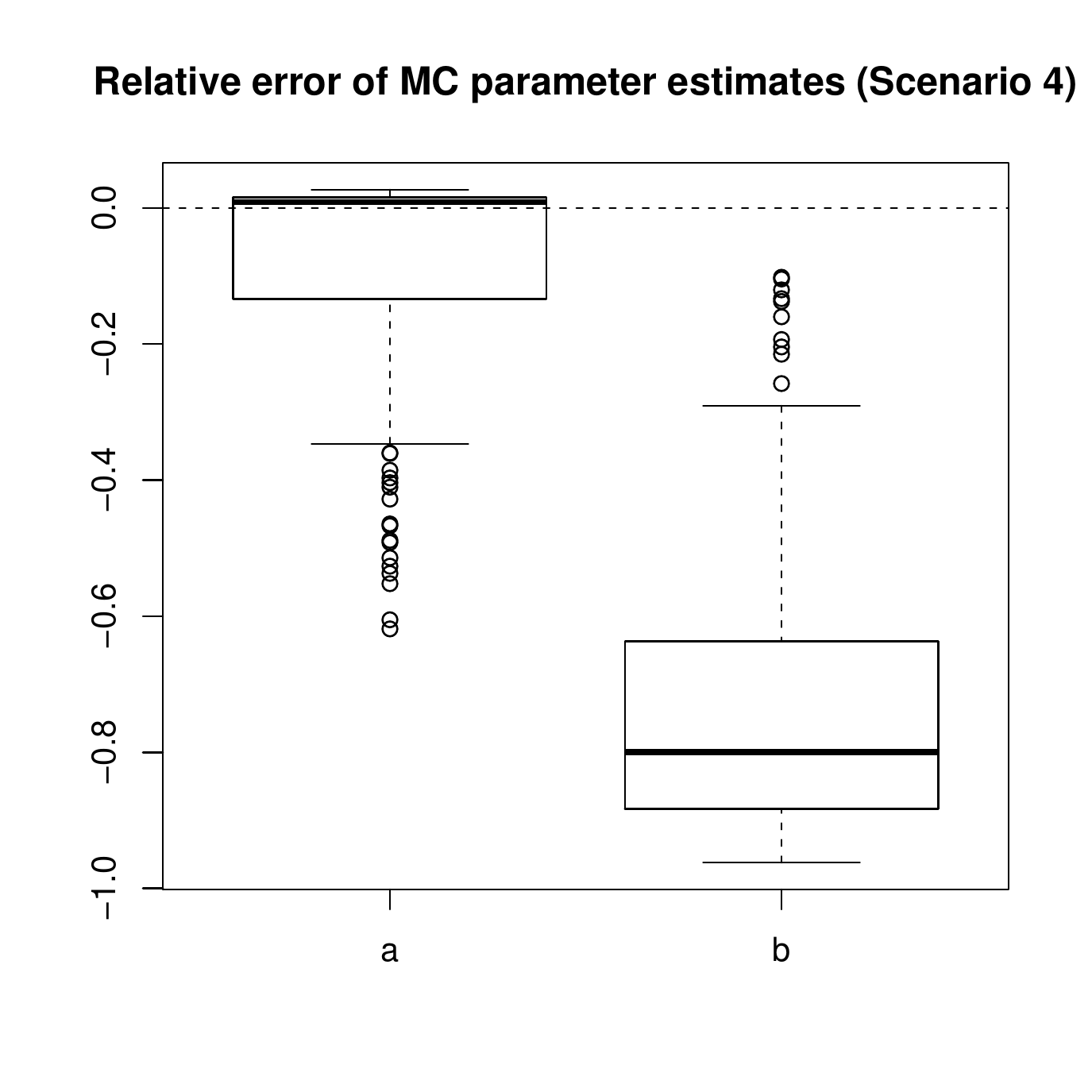}\vspace{-.8cm}

\begin{tabular}{c c  c}
MC parameters &$a$ & $b$ \\
\hline
relative bias  & -9.7 $\cdot 10^{-2}$ & -7.2 $\cdot 10^{-1}$  \\
\hline
relative MSE   & 4.0 $\cdot 10^{-2} $ & 5.1 $ \cdot 10^{-1}$  \\
\end{tabular}
\end{minipage}
\captionof{table}{Relative error of Markov chain parameter estimates in Scenarios 2 (left figure) and 4 (right figure), and relative bias / MSE for Scenario 2 (left table) and  Scenario 4 (right table).}
\label{simstudy_mcbias}
\end{minipage}

\begin{minipage}{\textwidth}
\includegraphics[width=0.9\textwidth]{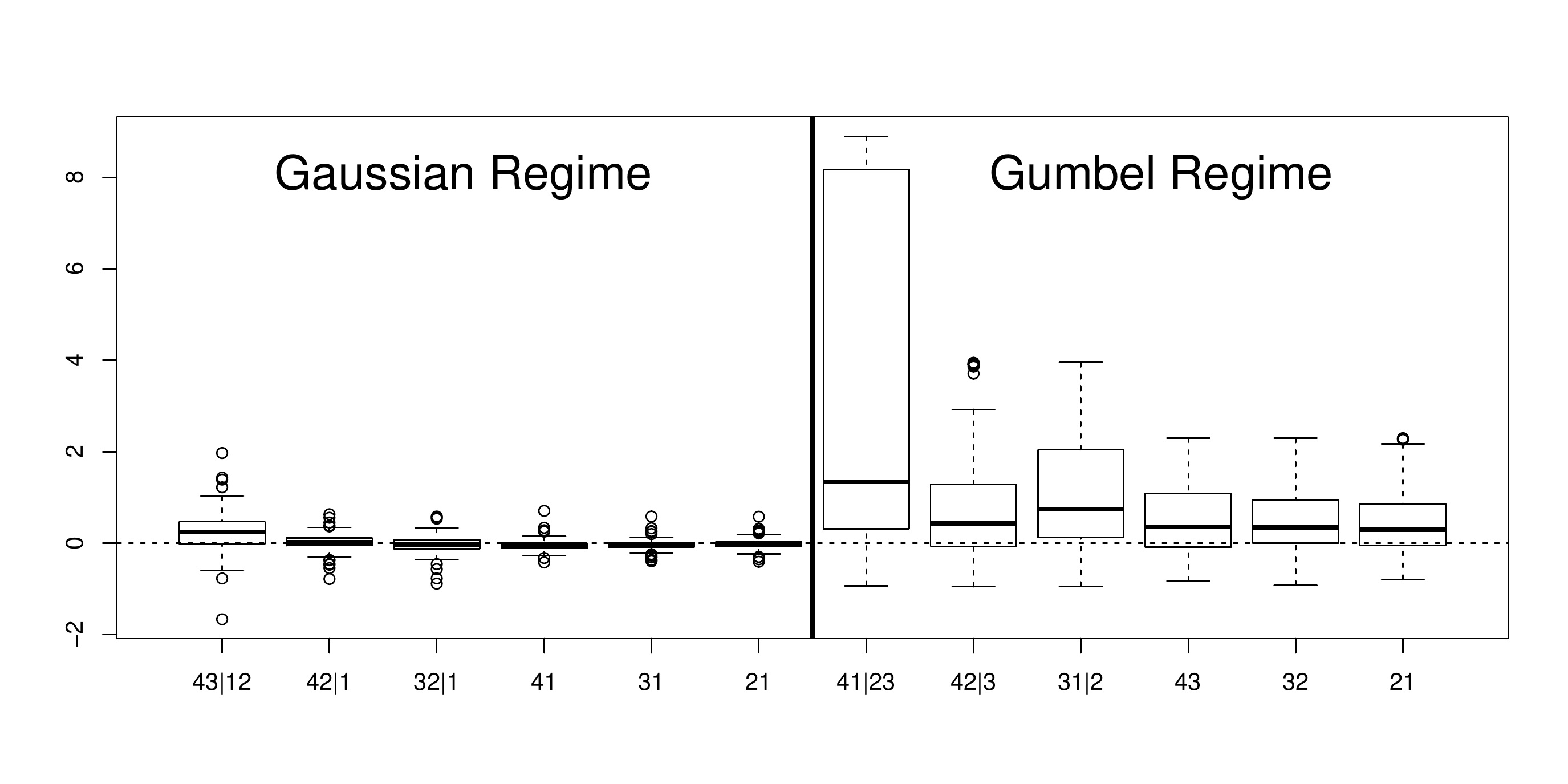}
\centering
\begin{tabular}{c  c  c  c  c  c  c }
Gaussian Regime & $\tau_{43 \vert 12}$ & $\tau_{42 \vert 1}$ & $\tau_{32 \vert 1}$ & $\tau_{41}$ & $\tau_{31}$ &$\tau_{21}$ \\
\hline
relative bias  & 2.4 $\cdot 10^{-1}$ & 1.9 $\cdot 10^{-2}$ & -2.6 $\cdot 10^{-2}$ & -5.1 $\cdot 10^{-2}$ & -3.7 $\cdot 10^{-2}$ & -1.5 $\cdot 10^{-2}$ \\
relative MSE   & 2.5 $\cdot 10^{-2} $ & 7.5 $ \cdot 10^{-3}$ & 8.3 $\cdot 10^{-3}$ & 5.7 $\cdot 10^{-3}$ & 5.4 $ \cdot 10^{-3} $ & 4.4 $\cdot 10^{-3} $ \\
\hline
\hline
Gumbel Regime & $\tau_{41 \vert 23}$ & $\tau_{42 \vert 3}$ & $\tau_{31 \vert 2}$ & $\tau_{43}$ & $\tau_{32}$ &$\tau_{21}$ \\
\hline
relative bias  & 3.1 $\cdot 10^{0}$ & 8.7 $\cdot 10^{-1}$ & 1.2 $\cdot 10^{0}$ & 5.2 $\cdot 10^{1}$ & 5.1 $\cdot 10^{-1}$ & 4.5 $\cdot 10^{-1}$ \\
relative MSE   & 2.3 $\cdot 10^{0} $ & 5.3 $ \cdot 10^{-1}$ & 7.5 $\cdot 10^{-1}$ & 2.7 $\cdot 10^{-1}$ & 2.3 $ \cdot 10^{-1} $ & 2.1 $\cdot 10^{-1} $ \\
\end{tabular}
\captionof{table}{Scenario 4: Relative error of Kendall's $\tau$ estimates (top figure) and relative bias / MSE for the Gaussian regime (upper table) and the Gumbel regime (lower table), respectively.}
\end{minipage}

\section{Selected R-vine structures}\label{appendix_rvine_structure}
\subsection{Selected R-vine structure for the US exchange rates}
\subsubsection*{Model (1)}

In Section \ref{ex_Rvine} an R-vine is fitted to the exchange rate dataset using the outlined procedures of \shortciteN{dissmann2011} and \shortciteN{brechmann2011b}. Here only the copula parameters are switching while tree structure and copula families are common to both regimes. The resulting tree structure is given in Figure \ref{RVineTreePlot}
the following together with corresponding Kendall's $\tau$ estimates in Table \ref{param_model1}.

\pagebreak
\begin{minipage}{\textwidth}
\centering
\begin{tabular}{c  c  c  c  c  c  c  c  c }
Tree 1 & {\footnotesize GBP,EUR} & {\footnotesize EUR,CHF} & {\footnotesize CHF,JPY} & {\footnotesize AUS,EUR} & {\footnotesize AUS,BRL} & {\footnotesize INR,AUS} & {\footnotesize CAD,AUS} & {\footnotesize CNY,INR} \\
\hline
cop. fam. & SG & SG & SG & N & G & N & N & G \\
\hline
\hline
\multicolumn{9}{c}{Regime 1}\\
\hline
$\hat \btau_1^{EM}$  		& 0.55 & 0.78 & 0.46 & 0.46 & 0.19 & 0.14 & 0.29 & 0.11 \\
$\hat \btau_1^{MCMC}$ 		& 0.56 & 0.79 & 0.47 & 0.46 & 0.18 & 0.14 & 0.28 & 0.13 \\
5\% quant.				& 0.53 & 0.77 & 0.43 & 0.43 & 0.14 & 0.10 & 0.24 & 0.09 \\
95\% quant.  					& 0.60 & 0.81 & 0.50 & 0.49 & 0.23 & 0.20 & 0.32 & 0.17 \\
\hline 
\hline
\multicolumn{9}{c}{Regime 2}\\
\hline
$\hat \btau_2^{EM}$  		& 0.44 & 0.58 & 0.24 & 0.41 & 0.45 & 0.26 & 0.44 & 0.07 \\
$\hat \btau_2^{MCMC}$ 		& 0.44 & 0.58 & 0.22 & 0.40 & 0.43 & 0.25 & 0.44 & 0.05 \\
5\% quant. 				& 0.40 & 0.55 & 0.17 & 0.36 & 0.39 & 0.20 & 0.40 & 0.02 \\
95\% quant. 						& 0.47 & 0.60 & 0.27 & 0.43 & 0.47 & 0.29 & 0.50 & 0.10 \\
\end{tabular}
\vspace{.5cm}

\begin{tabular}{c  c  c  c  c  c  c  c }
\multirow{2}{*}{Tree 2} & {\footnotesize GBP,AUS$\vert$} & {\footnotesize CAD,EUR$\vert$} & {\footnotesize CAD,BRL$\vert$} & {\footnotesize BRL,INR$\vert$} & {\footnotesize CNY,AUS$\vert$} & {\footnotesize JPY,EUR$\vert$} &   {\footnotesize CHF,AUS$\vert$}  \\
&{\footnotesize EUR} & {\footnotesize AUS} & {\footnotesize AUS} & {\footnotesize AUS} & {\footnotesize INR} & {\footnotesize CHF} & {\footnotesize EUR} \\
\hline
cop. fam. & G & G & SG & 0 & SG & G270 & G270  \\
\hline
\hline
\multicolumn{8}{c}{Regime 1}\\
\hline
$\hat \btau_1^{EM}$  	& 0.15 & 0.11 & 0.07 & 0.02 & 0.01 & -0.06 & -0.03  \\
$\hat \btau_1^{MCMC}$ 	& 0.15 & 0.11 & 0.07 & 0.01 & 0.02 & -0.06 & -0.03  \\
5\% quant. 			& 0.10 & 0.07 & 0.03 & -0.04 & 0.00 & -0.10 & -0.07  \\
95\% quant. 				& 0.20 & 0.15 & 0.13 & 0.06 & 0.05 & -0.02 & -0.00  \\
\hline 
\hline
\multicolumn{8}{c}{Regime 2}\\
\hline
$\hat \btau_2^{EM}$  	& 0.15 & 0.11 & 0.11 & 0.11 & 0.10 & -0.31 & -0.24  \\
$\hat \btau_2^{MCMC}$ 	& 0.16 & 0.13 & 0.11 & 0.11 & 0.11 & -0.31 & -0.24  \\
5\% quant.			& 0.10 & 0.08 & 0.04 & 0.06 & 0.07 & -0.36 & -0.28  \\
95\% quant. 				& 0.21 & 0.17 & 0.17 & 0.16 & 0.16 & -0.26 & -0.19  \\
\end{tabular}

\captionof{table}{Estimated Kendall's $\tau$ for the first and second tree of Model (1).}
\label{param_model1}
\end{minipage}

\begin{minipage}{\textwidth}
\begin{minipage}{0.5\textwidth}
\includegraphics[width=\textwidth]{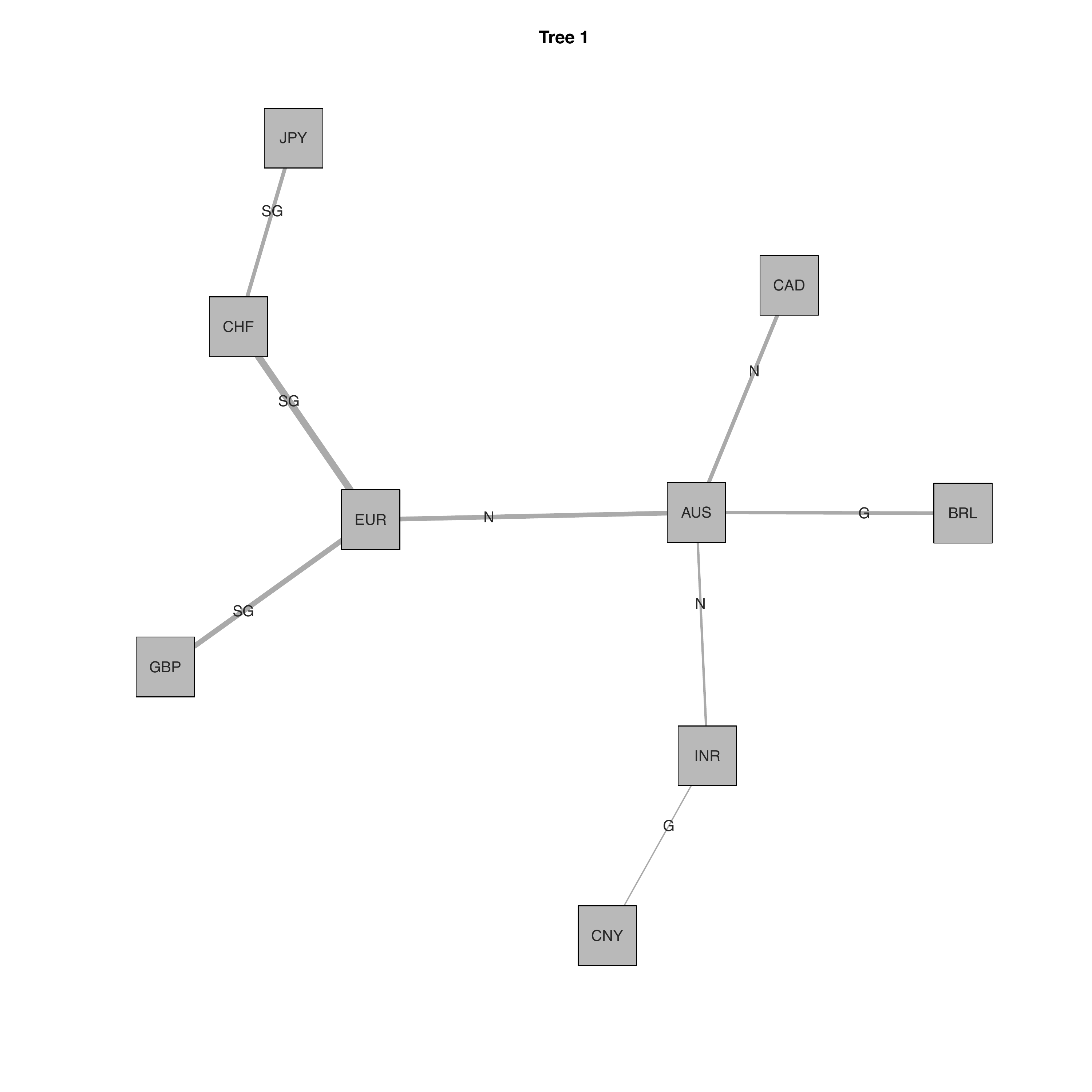}
\end{minipage}
\begin{minipage}{0.5\textwidth}
\includegraphics[width=\textwidth]{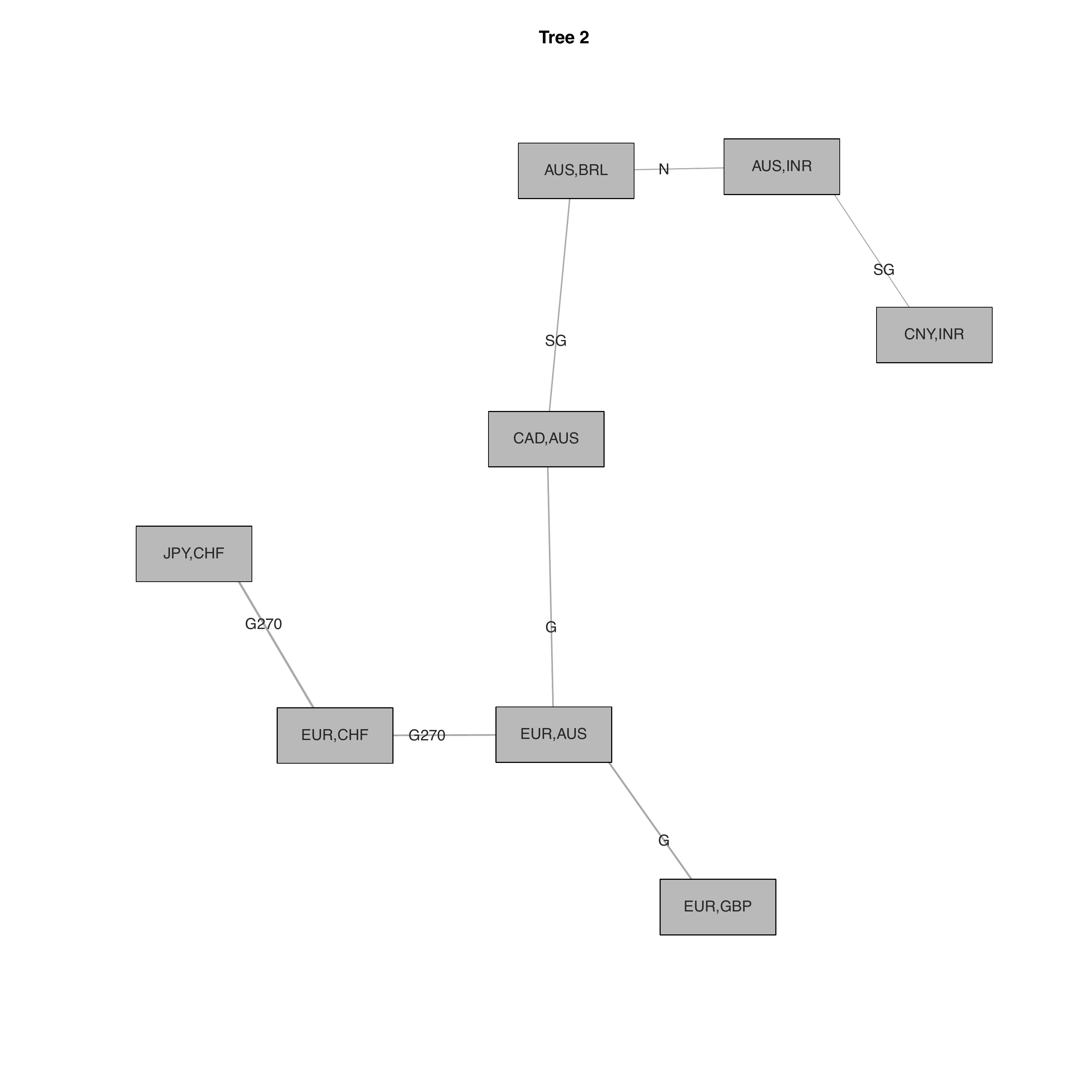}
\end{minipage}
\captionof{figure}{The first and second tree of the R-vine tree structure $\mathcal{V}_1$ of Model (1) for the exchange rate data. We also choose this structure for the non-crisis regime in Models (2a) - (2c) and Model (3). }
\label{RVineTreePlot}
\end{minipage}

\subsubsection*{Model (2)}
For Models (2a) - (2c), two R-vine tree structures have been selected to account for different dependencies during times of crisis and normal times. The first R-vine ($\mathcal{V}_1$), corresponding to normal times, has again the structure displayed in Figure \ref{RVineTreePlot}, the first and second tree of the second R-vine ($\mathcal{V}_2$), corresponding to crisis times, is given in Figure \ref{ex_crisis_rvines}.

\begin{minipage}{\textwidth}
\begin{minipage}{0.5\textwidth}
\includegraphics[width=\textwidth]{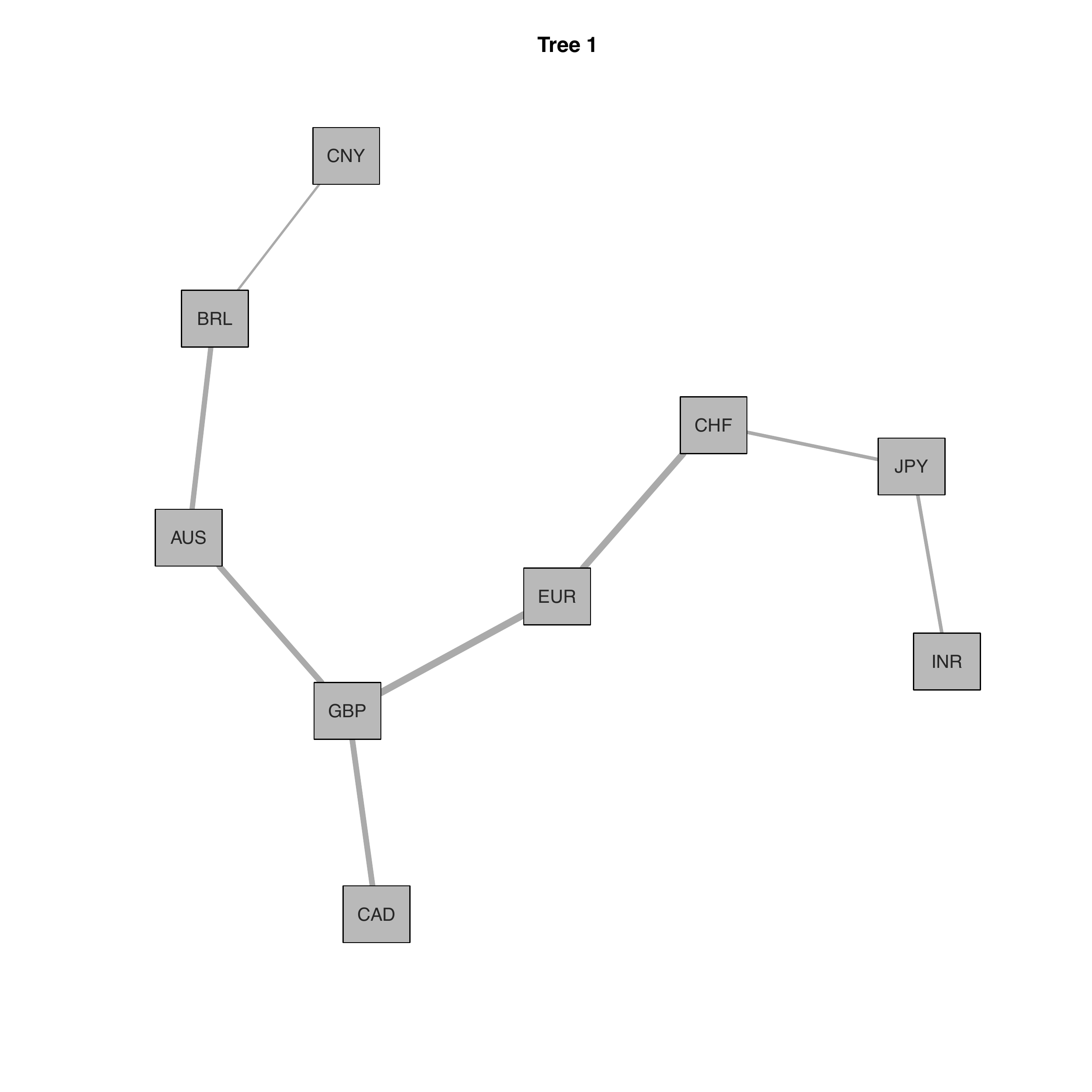}
\end{minipage}
\begin{minipage}{0.5\textwidth}
\includegraphics[width=\textwidth]{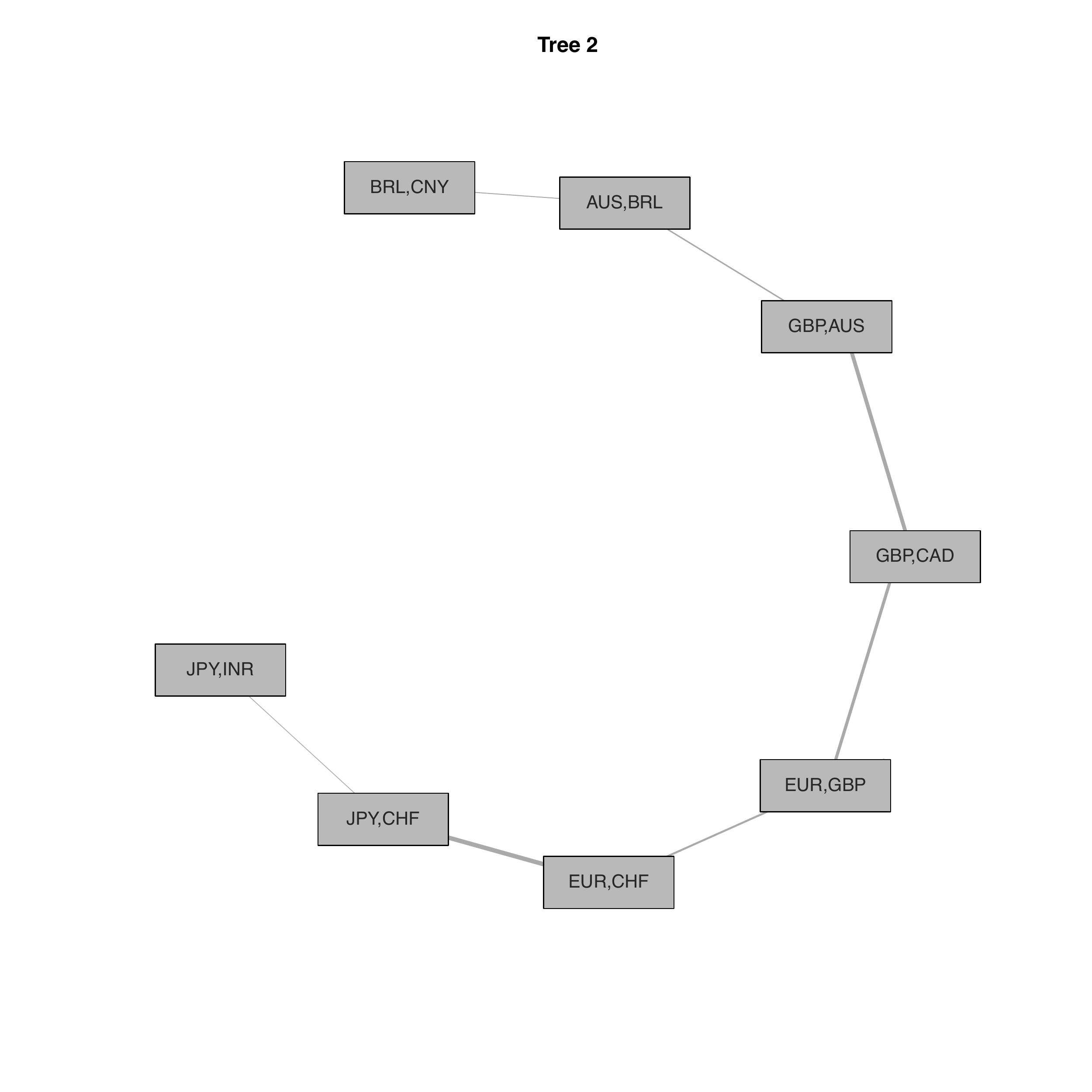}
\end{minipage}\vspace{-1cm}
\captionof{figure}{The first and second tree of the "crisis" R-vine structure $\mathcal{V}_2$ which we have chosen for Model (2) of the US exchange rate data.}
\label{ex_crisis_rvines}
\end{minipage}

{\small
\begin{minipage}{\textwidth}
\centering
\begin{tabular}{c c  c  c  c  c  c c  c }
"normal", $\mathcal{V}_1$ & {\footnotesize GBP,EUR} & {\footnotesize EUR,CHF} & {\footnotesize CHF,JPY} & {\footnotesize AUS,EUR} & {\footnotesize AUS,BRL} & {\footnotesize INR,AUS} & {\footnotesize  CAD,AUS} & {\footnotesize CNY,INR} \\
\hline
(2a) $\hat \btau_1^{EM}$  	& 0.53 & 0.75 & 0.45 & 0.46 & 0.28 & 0.21 & 0.35 & 0.12 \\
(2a) $\hat \btau_1^{MCMC}$ 	& 0.56 & 0.78 & 0.45 & 0.48 & 0.24 & 0.21 & 0.33 & 0.15 \\
5\% quant.				& 0.52 & 0.75 & 0.41 & 0.45 & 0.19 & 0.17 & 0.29 & 0.10 \\
95\% quant. 					& 0.60 & 0.80 & 0.49 & 0.51 & 0.29 & 0.25 & 0.36 & 0.20 \\
\hline 
\hline
(2b) $\hat \btau_1^{EM}$  	& 0.54 & 0.75 & 0.44 & 0.47 & 0.29 & 0.22 & 0.35 & 0.12 \\
(2b) $\hat \btau_1^{MCMC}$ 	& 0.52 & 0.74 & 0.43 & 0.44 & 0.29 & 0.21 & 0.34 & 0.11 \\
5\% quant.  				& 0.49 & 0.72 & 0.40 & 0.41 & 0.26 & 0.17 & 0.31 & 0.07 \\
95\% quant.				& 0.55 & 0.76 & 0.46 & 0.48 & 0.32 & 0.24 & 0.37 & 0.14 \\
\hline 
\hline
(2c) $\hat \btau_1^{EM}$  		& 0.60 & 0.81 & 0.47 & 0.48 & 0.21 & 0.19 & 0.32 & 0.16 \\
(2c) $\hat \btau_1^{MCMC}$ 	& 0.60 & 0.80 & 0.47 & 0.49 & 0.21 & 0.21 & 0.31 & 0.17 \\
5\% quant. 				& 0.57 & 0.79 & 0.44 & 0.46 & 0.16 & 0.17 & 0.27 & 0.13 \\
95\% quant. 					& 0.62 & 0.82 & 0.51 & 0.52 & 0.26 & 0.22 & 0.35 & 0.22 \\
\end{tabular}
\vspace{.3cm}

\begin{tabular}{c c  c  c  c  c  c c  c }
"crisis", $\mathcal{V}_2$  & {\footnotesize GBP,EUR }&  {\footnotesize EUR,CHF }&  {\footnotesize CHF,JPY }&  {\footnotesize JPY,INR }&  {\footnotesize AUS,GBP }&  {\footnotesize BRL,AUS }&  {\footnotesize  BRL,CNY }&  {\footnotesize CAD,GBP} \\
\hline
(2a) $\hat \btau_2^{EM}$        		& 0.44 & 0.45 & 0.11 & 0.00 & 0.41 & 0.49 & 0.11 & 0.41 \\
(2a) $\hat \btau_2^{MCMC}$ 		& 0.42 & 0.52 & 0.22 & 0.01 & 0.37 & 0.47 & 0.07 & 0.37 \\
5\% quant. 					& 0.36 & 0.47 & 0.11 & 0.00 & 0.30 & 0.41 & 0.01 & 0.29 \\
95\% quant.						& 0.49 & 0.56 & 0.30 & 0.02 & 0.44 & 0.53 & 0.13 & 0.43 \\
\hline 
\hline
(2b) $\hat \btau_2^{EM}$  		& 0.37 & 0.37 & 0.10 & 0.00 & 0.32 & 0.41 & 0.08 & 0.36 \\
(2b) $\hat \btau_2^{MCMC}$ 		& 0.45 & 0.37 & 0.05 & 0.01 & 0.35 & 0.44 & 0.13 & 0.38 \\
5\% quant. 					& 0.34 & 0.27 & 0.00 & 0.00 & 0.25 & 0.36 & 0.03 & 0.30 \\
95\% quant.						& 0.55 & 0.45 & 0.15 & 0.04 & 0.44 & 0.53 & 0.24 & 0.46 \\
\hline 
\hline
(2c) $\hat \btau_2^{EM}$  			&  {\footnotesize 0.43, 10.8 }&  {\footnotesize 0.58, 8.6 }&  {\footnotesize 0.27, 7.9}&  {\footnotesize -0.13, 30 }&  {\footnotesize 0.37, 10.7 }&  {\footnotesize 0.44, 5.9 }&  {\footnotesize 0.06, 30 }&  {\footnotesize 0.33, 30} \\
(2c) $\hat \btau_2^{MCMC}$ 		& {\footnotesize 0.42, 14.2 }&  {\footnotesize 0.56, 9.7 }&  {\footnotesize 0.25, 9.8 }&  {\footnotesize -0.15, 21.4 }&  {\footnotesize 0.35, 15.4 }&  {\footnotesize 0.45, 9.3 }&  {\footnotesize 0.05, 20.8 }&  {\footnotesize 0.34, 21.8} \\
5\% quant.  					& {\footnotesize 0.38, 7.0}&  {\footnotesize 0.53, 5.6 }&  {\footnotesize 0.21, 5.2 }&  {\footnotesize -0.21, 11.6 }&  {\footnotesize 0.31, 7.0 }&  {\footnotesize 0.45, 4.8 }&  {\footnotesize -0.01, 10.7 }&  {\footnotesize 0.29, 11.0} \\
95 \% quant.						&  {\footnotesize 0.46, 25.7 }&  {\footnotesize 0.60, 16.3 }&  {\footnotesize 0.29, 18.3 }&  {\footnotesize -0.11, 29.2 }&  {\footnotesize 0.40, 28.1 }&  {\footnotesize 0.49, 19.0 }&  {\footnotesize 0.11, 29.2} &  {\footnotesize 0.39, 29.3} \\
\end{tabular}
\captionof{table}{Estimated Kendall's $\tau$ values corresponding to the first tree of Models (2a) - (2c), respectively. For the t-copula used in Model (2c), the first parameter is transformed to Kendall's $\tau$, the second parameter gives the estimated degrees of freedom (with $\nu=30$ as upper limit.}
\label{param_model2_1}
\end{minipage}
}

{\small
\begin{minipage}{1.\textwidth}
\centering
\begin{tabular}{c  c  c  c  c  c  c  c  }
\multirow{2}{*}{"normal", $\mathcal{V}_1$ } &  {\footnotesize JPY,EUR$\vert$} &  {\footnotesize AUS,CHF$\vert$ }&  {\footnotesize AUS,GBP$\vert$ }&  {\footnotesize CAD,EUR$\vert$ }&  {\footnotesize CAD,BRL$\vert$ }&  {\footnotesize INR,BRL$\vert$ }&  {\footnotesize  CNY,AUS$\vert$} \\
& {\footnotesize CHF }&  {\footnotesize EUR }&  {\footnotesize EUR }&  {\footnotesize AUS} &  {\footnotesize AUS} &  {\footnotesize AUS} &  {\footnotesize INR}\\
\hline
(2a) $\hat \btau_1^{EM}$  	& -0.14 & -0.13 & 0.14 & 0.10 & 0.12 & 0.07 & 0.01  \\
(2a) $\hat \btau_1^{MCMC}$ 	& -0.10 & -0.08 & 0.13 & 0.10 & 0.11 & 0.06 & 0.00  \\
5 \% quantile			& -0.16 & -0.14 & 0.09 & 0.06 & 0.07 & 0.01 & -0.05  \\
95 \% quantile 					& -0.02 & 0.00 & 0.18 & 0.14 & 0.15 & 0.10 & 0.04  \\
\hline 
\hline
(2b) $\hat \btau_1^{EM}$   	& -0.16 & -0.14 & 0.15 & 0.10 & 0.12 & 0.07 & 0.01  \\
(2b) $\hat \btau_1^{MCMC}$ 	& -0.17 & -0.17 & 0.16 & 0.10 & 0.11 & 0.07 & 0.02 \\
5 \% quantile			& -0.21 & -0.20 & 0.13 & 0.06 & 0.07 & 0.03 & -0.02  \\
95 \% quantile 				 	& -0.13 & -0.13 & 0.20 & 0.13 & 0.14 & 0.11 & 0.06  \\
\hline 
\hline
(2c) $\hat \btau_1^{EM}$  		& -0.01 & 0.02 & 0.15 & 0.10 & 0.12 & 0.03 & 0.00  \\
(2c) $\hat \btau_1^{MCMC}$ 	& -0.03 & 0.00 & 0.14 & 0.10 & 0.11 & 0.04 & 0.00  \\
5 \% quantile			& -0.08 & -0.06 & 0.09 & 0.05 & 0.07 & -0.01 & -0.05  \\
95 \% quantile 					&  0.03 & 0.05 & 0.18 & 0.15 & 0.16 & 0.09 & 0.04  \\
\end{tabular}
\vspace{0.2cm}

\begin{tabular}{c  c c  c  c  c  c  c  }
\multirow{2}{*}{"crisis", $\mathcal{V}_2$ } & {\footnotesize CNY,AUS$\vert$ }&  {\footnotesize GBP,BRL$\vert$ }&  {\footnotesize AUS,CAD$\vert$ }&  {\footnotesize CAD,EUR$\vert$ }&  {\footnotesize GBP,CHF$\vert$ }&  {\footnotesize EUR,JPY$\vert$ }&  {\footnotesize  CHF,INR$\vert$} \\
&  {\footnotesize BRL }&  {\footnotesize AUS }&  {\footnotesize GBP }&  {\footnotesize GBP }&  {\footnotesize EUR }&  {\footnotesize CHF }&  {\footnotesize JPY}\\
\hline
(2a) $\hat \btau_2^{EM}$  	& 0.21 & 0.04 & 0.24 & 0.14 & -0.22 & -0.42 & 0.03  \\
(2a) $\hat \btau_2^{MCMC}$ 	& 0.17 & 0.04 & 0.28 & 0.15 & -0.19 & -0.36 & 0.07  \\
5 \% quantile			& 0.09 & -0.03 & 0.19 & 0.07 & -0.26 & -0.45 & 0.01  \\
95 \% quantile					& 0.25 & 0.12 & 0.35 & 0.21 & -0.13 & -0.31 & 0.15 \\
\hline 
\hline
(2b) $\hat \btau_2^{EM}$  	& 0.19 & 0.11 & 0.26 & 0.17 & -0.16 & -0.36 & -0.02  \\
(2b) $\hat \btau_2^{MCMC}$ 	& 0.22 & 0.14 & 0.39 & 0.28 & -0.14 & -0.42 & -0.02  \\
5 \% quantile				& 0.11 & 0.02 & 0.21 & 0.14 & -0.23 & -0.52 & -0.10  \\
95 \% quantile				& 0.34 & 0.28 & 0.56 & 0.42 & -0.04 & -0.31 & 0.07   \\
\hline 
\hline
(2c) $\hat \btau_2^{EM}$  		& 0.10 & 0.03 & 0.30 & 0.18 & -0.15 & -0.35 & 0.12  \\
(2c) $\hat \btau_2^{MCMC}$ 	& 0.11 & 0.04 & 0.30 & 0.18 & -0.16 & -0.36 & 0.11  \\
5 \% quantile				& 0.06 & -0.01 & 0.25 & 0.12 & -0.20 & -0.40 & 0.05  \\
95 \% quantile				& 0.16 & 0.10 & 0.35 & 0.22 & -0.11 & -0.31 & 0.16  \\
\end{tabular}
\captionof{table}{Estimated Kendall's $\tau$, corresponding to the second tree of Models (2a) - (2c).}
\label{param_model2_2}
\end{minipage}
}

\subsubsection*{Model (3)}
Also for Model (3), we consider two different R-vine structures too account for changes in dependence. The structure for the first regime is again chosen to be $\mathcal{V}_1$, with copulas selected by AIC, the tree structure for the second regime is $\mathcal{V}_3$, selected from the part of the data labeled "crisis" in Figure \ref{rolling_window_ll} and shown in Figure \ref{RVineTreePlot3}. The corresponding copulas, together with estimated values for Kendall's $\tau$, are given in Table \ref{model3table1} for the "normal" regime and in Table \ref{model3table2} for the "crisis" regime, respectively.

\begin{minipage}{\textwidth}
\begin{minipage}{0.5\textwidth}
\includegraphics[width=\textwidth]{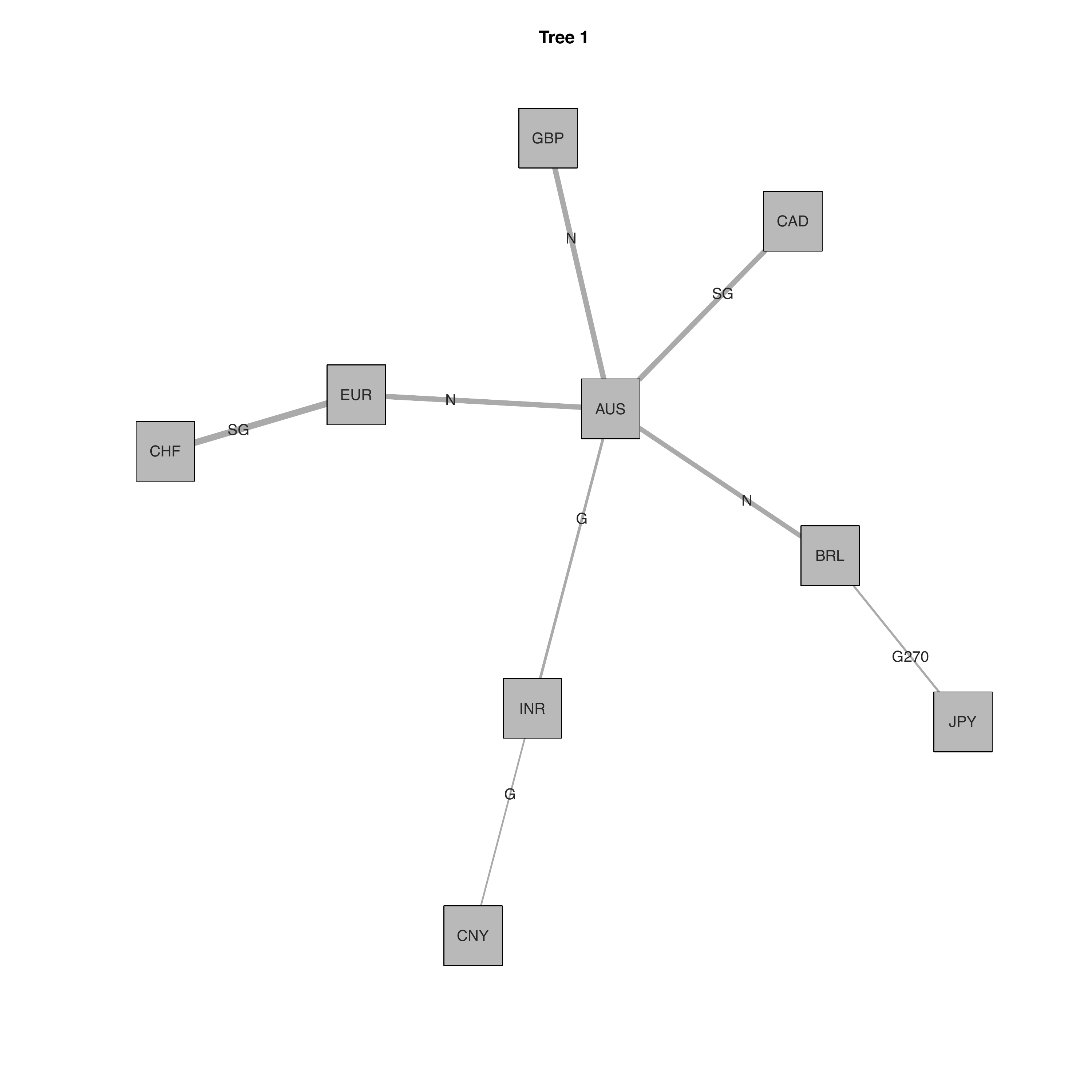}
\end{minipage}
\begin{minipage}{0.5\textwidth}
\includegraphics[width=\textwidth]{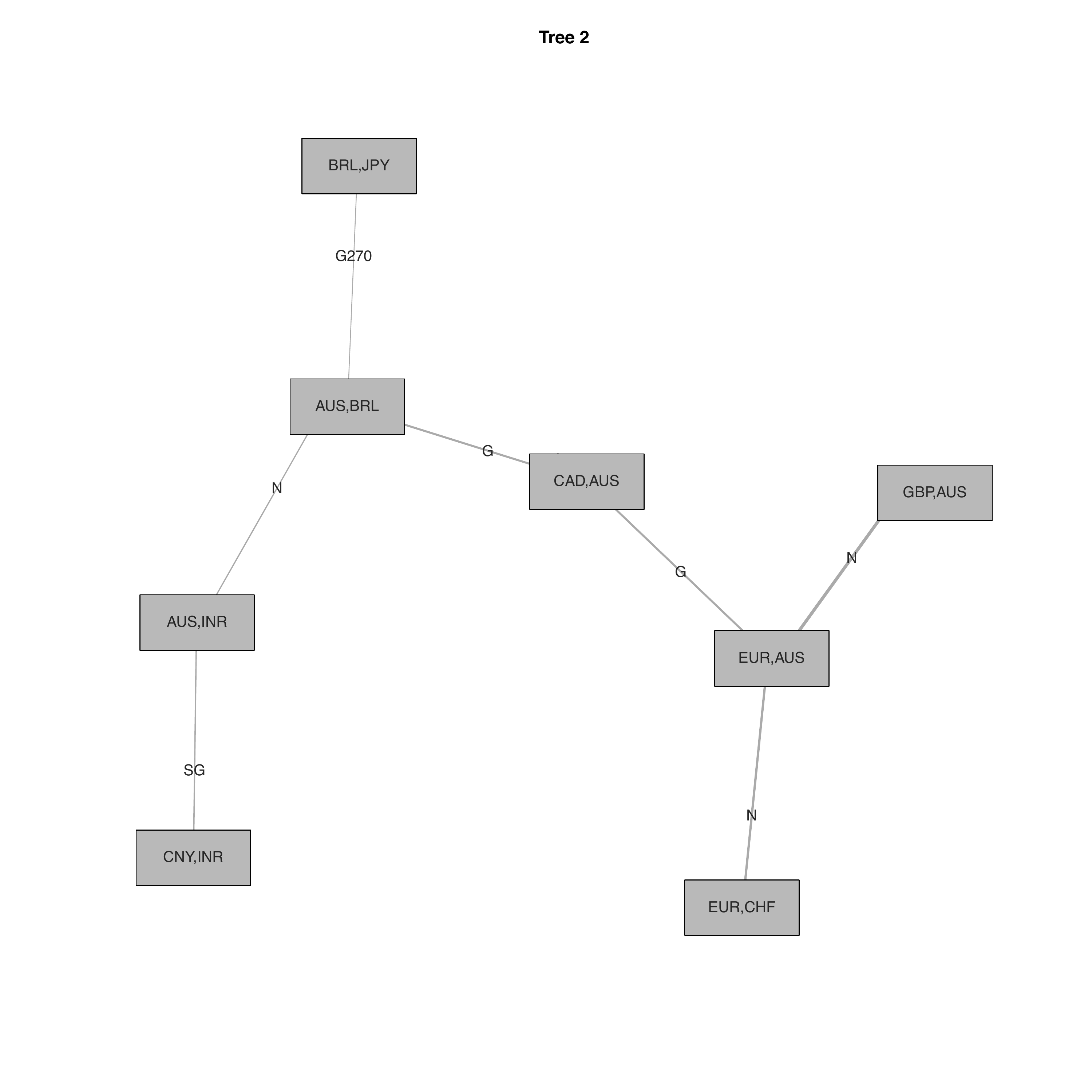}
\end{minipage}
\captionof{figure}{The first and second tree of the R-vine structure which we have chosen for the "crisis" regime of Model (3). We refer to this structure as $\mathcal{V}_3$.}
\label{RVineTreePlot3}
\end{minipage}

{\small
\begin{minipage}{\textwidth}
\centering
\vspace{.3cm}
\begin{tabular}{c  c  c  c  c  c  c  c  c }
"normal", $\mathcal{V}_1$ & {\footnotesize GBP,EUR }& {\footnotesize EUR,CHF }& {\footnotesize CHF,JPY }& {\footnotesize AUS,EUR }& {\footnotesize AUS,BRL }& {\footnotesize INR,AUS }& {\footnotesize  CAD,AUS }& {\footnotesize CNY,INR} \\
\hline
cop. fam. & SG & N & N & N & G & N & N & G \\
\hline
(2a) $\hat \btau_1^{EM}$  	& 0.51 & 0.76 & 0.49 & 0.45 & 0.22 & 0.16 & 0.30 & 0.11 \\
(2a) $\hat \btau_1^{MCMC}$ 	& 0.52 & 0.76 & 0.49 & 0.45 & 0.22 & 0.16 & 0.30 & 0.11 \\
5 \% quantile				& 0.49 & 0.75 & 0.46 & 0.42 & 0.17 & 0.12 & 0.26 & 0.07 \\
95 \% quantile				& 0.55 & 0.78 & 0.52 & 0.48 & 0.26 & 0.20 & 0.34 & 0.16 \\
\end{tabular}

\vspace{.3cm}

\begin{tabular}{c  c  c  c  c  c  c  c  }
\multirow{2}{*}{"normal", $\mathcal{V}_1$ } & {\footnotesize JPY,EUR$\vert$ }& {\footnotesize AUS,CHF$\vert$ }& {\footnotesize AUS,GBP$\vert$ }& {\footnotesize CAD,EUR$\vert$ }& {\footnotesize CAD,BRL$\vert$ }& {\footnotesize INR,BRL$\vert$ }& {\footnotesize  CNY,AUS$\vert$} \\
& {\footnotesize CHF }& {\footnotesize EUR }& {\footnotesize EUR }& {\footnotesize AUS }& {\footnotesize AUS }& {\footnotesize AUS }& {\footnotesize INR}\\
\hline
cop. fam. & G270 & G 270 & G & G & N & N & G  \\
\hline
(2a) $\hat \btau_1^{EM}$  	& -0.09 & -0.05 & 0.14 & 0.10 & 0.08 & 0.05 & 0.03  \\
(2a) $\hat \btau_1^{MCMC}$ 	& -0.09 & -0.05 & 0.14 & 0.11 & 0.08 & 0.04 & 0.04  \\
5\% quant.				& -0.14 & -0.10 & 0.10 & 0.07 & 0.04 & -0.01 & 0.01  \\
95\% quant.			& -0.04 & -0.02 & 0.18 & 0.15 & 0.13 & 0.09 & 0.07  \\
\end{tabular}
\captionof{table}{Estimated Kendall's $\tau$, corresponding to the first and second tree of the normal regime in Model (3).}
\label{model3table1}

\end{minipage}
}

{\small
\begin{minipage}{\textwidth}
\centering
\begin{tabular}{c  c  c  c  c  c  c  c  c }
"crisis", $\mathcal{V}_3$ & {\footnotesize CHF,EUR }& {\footnotesize EUR,AUS }& {\footnotesize GBP,AUS }& {\footnotesize AUS,CAD }& {\footnotesize AUS,BRL }& {\footnotesize BRL,JPY }& {\footnotesize  INR,AUS }& {\footnotesize CNY,INR} \\
\hline
cop. fam. & SG & N & N & SG & N & G270 & G & G \\
\hline
(2a) $\hat \btau_1^{EM}$  	& 0.54 & 0.42 & 0.42 & 0.50 & 0.52 & -0.34 & 0.23 & 0.06 \\
(2a) $\hat \btau_1^{MCMC}$ 	& 0.55 & 0.40 & 0.41 & 0.49 & 0.52 & -0.35 & 0.23 & 0.06 \\
5\% quant.				& 0.50 & 0.35 & 0.35 & 0.45 & 0.48 & -0.40 & 0.16 & 0.00 \\
95\% quant.					& 0.58 & 0.44 & 0.46 & 0.53 & 0.56 & -0.30 & 0.29 & 0.12 \\
\end{tabular}

\vspace{.2cm}

\begin{tabular}{c  c  c  c  c  c c  c  }
\multirow{2}{*}{"crisis", $\mathcal{V}_3$ } & {\footnotesize CNY,AUS$\vert$ }& {\footnotesize INR,BRL$\vert$ }& {\footnotesize AUS,JPY$\vert$ }& {\footnotesize CAD,BRL$\vert$ }& {\footnotesize CAD,EUR$\vert$ }& {\footnotesize  GBP,EUR$\vert$ }& {\footnotesize AUS,CHF$\vert$} \\
&{\footnotesize INR }& {\footnotesize AUS }& {\footnotesize BRL }& {\footnotesize AUS }& {\footnotesize AUS }& {\footnotesize AUS }& {\footnotesize EUR}\\
\hline
cop. fam. & SG & N & G270 & G & G & N  & N  \\
\hline
(2a) $\hat \btau_1^{EM}$  	& 0.10 & 0.11 & -0.14 & 0.10 & 0.15 & 0.34 & -0.32  \\
(2a) $\hat \btau_1^{MCMC}$ 	& 0.11 & 0.11 & -0.17 & 0.13 & 0.16 & 0.34 & -0.32  \\
5\% quant.				& 0.05 & 0.05 & -0.24 & 0.05 & 0.10 & 0.28 & -0.37  \\
95\% quant.				& 0.19 & 0.17 & -0.10 & 0.19 & 0.22 & 0.41 & -0.27  \\
\end{tabular}
\vspace{-.2cm}
\captionof{table}{Estimated Kendall's $\tau$, corresponding to the first and second tree of the crisis regime in Model (3).}
\label{model3table2}
\end{minipage}
}

\subsection{Selected R-vine structure for Eurozone country indices}
For the return data set of Eurozone country indices we selected a common R-vine structure (Figure \ref{stoxx_tree_figure}), with common copula families (Table \ref{stoxx_tree}) but MS parameters. The estimated Kendall's $\tau$ values are given in Table \ref{stoxx_tree}.

\vspace{.2cm}
\begin{minipage}{\textwidth}
\begin{minipage}{0.5\textwidth}
\includegraphics[width=.9\textwidth]{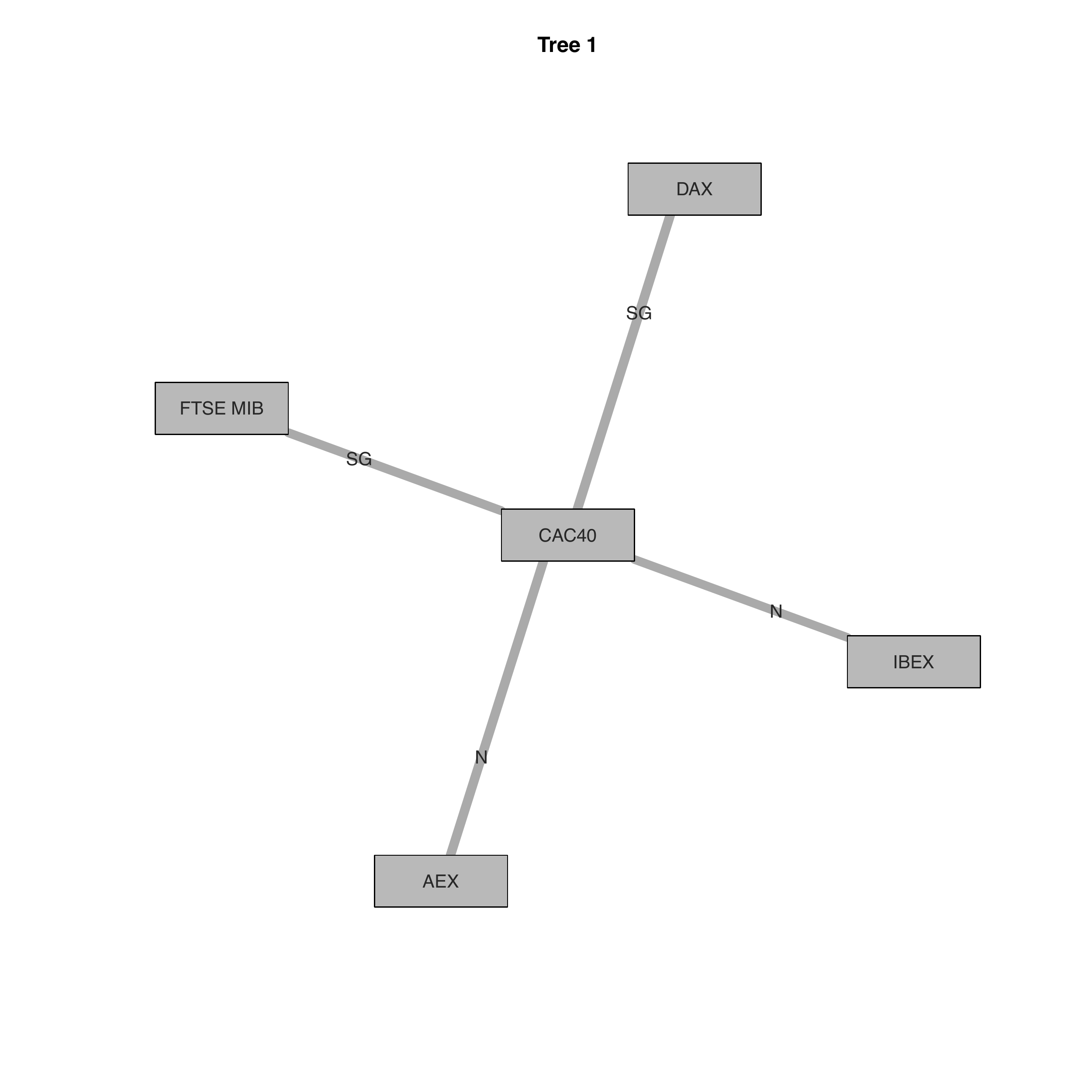}
\end{minipage}
\begin{minipage}{0.5\textwidth}
\includegraphics[width=.9\textwidth]{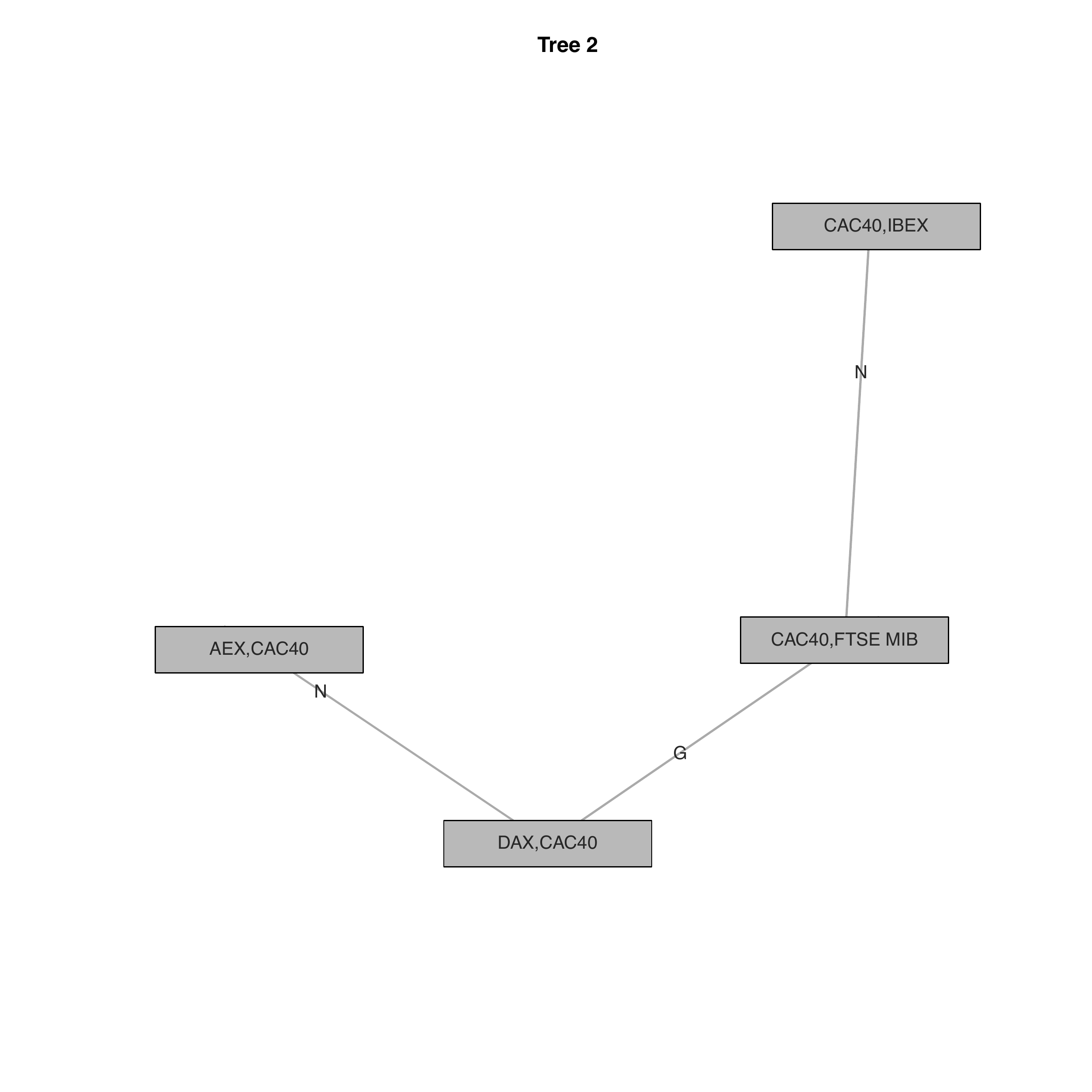}
\end{minipage}
\vspace{-1cm}
\captionof{figure}{First and second tree of the R-vine structure selected for Eurozone country indices.}
\label{stoxx_tree_figure}
\end{minipage}

\begin{minipage}{\textwidth}
\centering
{\small
\begin{tabular}{c  c  c  c c }
Tree 1 & DAX,CAC40 & IBEX,CAC40 & AEX,CAC40& FTSE MIB,CAC40 \\\hline
cop. fam. & SG & N & N & SG \\
$\hat \btau_1^{EM}${\scriptsize (Regime 1)} {\hspace{-.2cm}} & 0.71 & 0.62 & 0.72 & 0.66   \\
 $\hat \btau_2^{EM}${\scriptsize (Regime 2)} {\hspace{-.2cm}} & 0.83 & 0.78 & 0.81 & 0.77   \\
\end{tabular}
}
\vspace{.2cm}

{\small
\begin{tabular}{c c  c  c  }
\multirow{2}{*}{Tree 2} & AEX,DAX $\vert$ & DAX,FTSE MIB $\vert$ &  FTSE MIB,IBEX $\vert$  \\
& CAC40 & CAC40 & CAC40 \\
\hline
cop. fam. & N & G & N  \\
$\hat \btau_1^{EM}${\scriptsize (Regime 1)} {\hspace{-.2cm}}  & 0.06 & 0.13 & 0.12    \\
$\hat \btau_2^{EM}${\scriptsize (Regime 2)} {\hspace{-.2cm}} & 0.17 & 0.15 & 0.17    \\
\end{tabular}
}
\captionof{table}{Values of Kendall's $\tau$ estimated for the county indices.}
\label{stoxx_tree}
\end{minipage}

\subsection{Selected R-vine structure for the German stock index (DAX)}
\begin{minipage}{\textwidth}
\begin{minipage}{0.5\textwidth}
\includegraphics[width=.9\textwidth]{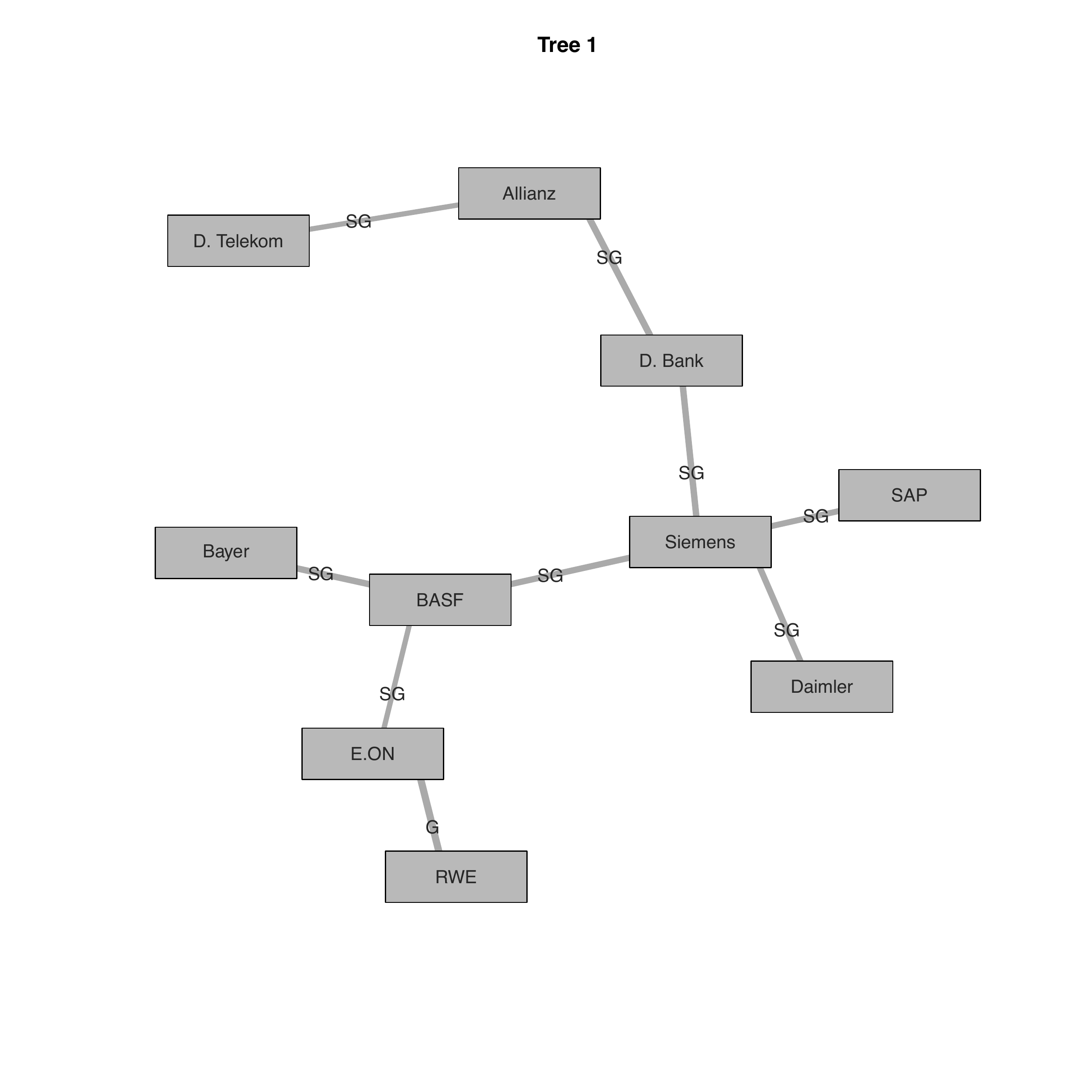}\end{minipage}
\begin{minipage}{0.5\textwidth}
\includegraphics[width=.9\textwidth]{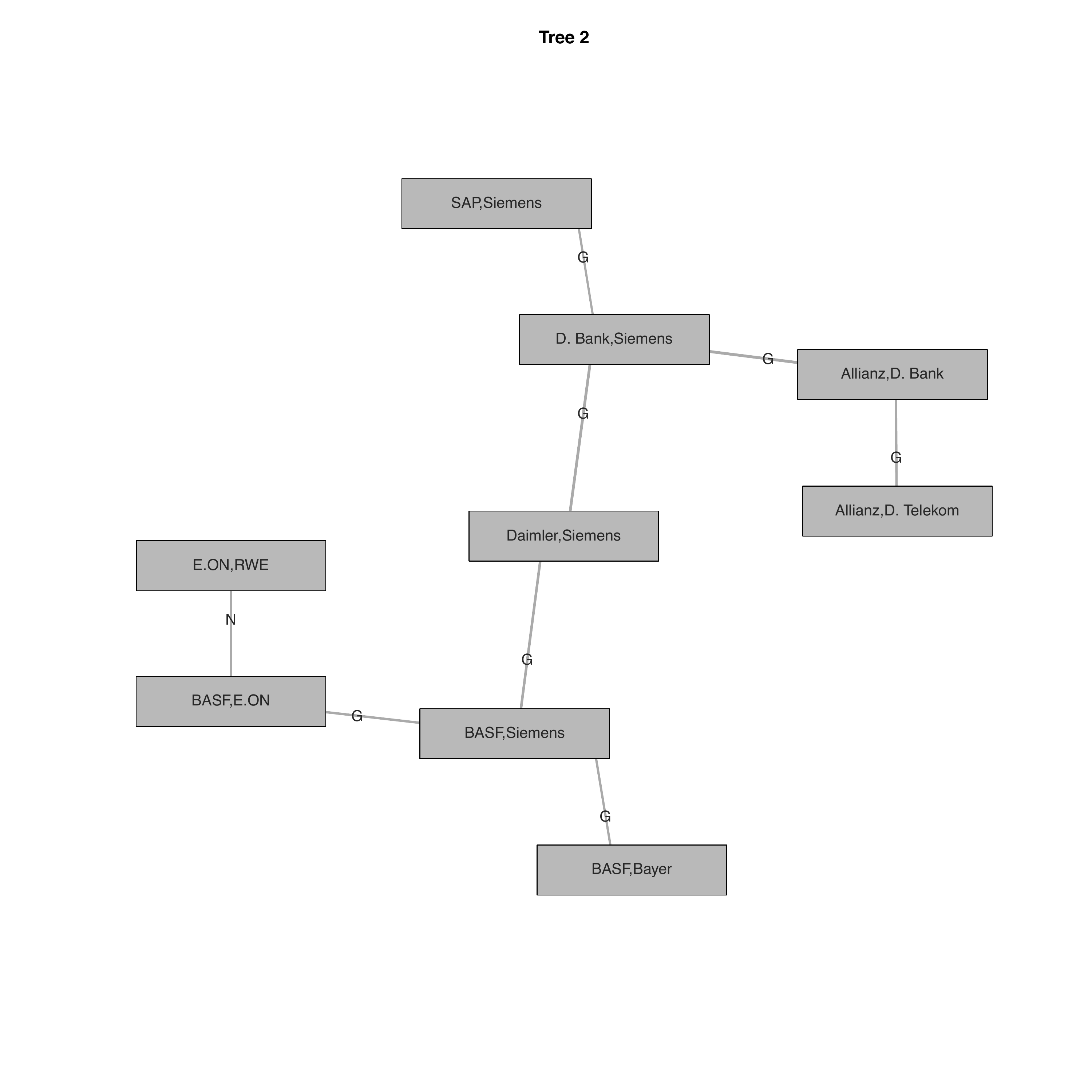}\end{minipage}
\vspace{-1cm}
\captionof{figure}{First and second tree of the R-vine structure selected for German stock returns. The corresponding values of Kendall's $\tau$ are given in Table \ref{dax_app_table}.}
\label{dax_trees_app}
\end{minipage}

{\small
\begin{minipage}{\textwidth}
\centering
\vspace{.2cm}
\begin{tabular}{c  c  c  c  c c  c  c  c  c  }

 Tree 1
 & {\footnotesize DTE,ALV}
 & {\footnotesize ALV,DBK}
 & {\footnotesize DBK,SIE}
  & {\footnotesize SIE,DAI}
 & {\footnotesize SIE,SAP}
 & {\footnotesize SIE,BAS}
 &  {\footnotesize BAS,BAY}
 & {\footnotesize BAS,EOA}
 &  {\footnotesize EOA,RWE} \\
\hline
cop. fam. & SG & SG & SG & SG & SG & SG & SG & SG & G \\

 $\hat \btau_1^{EM}${\scriptsize (Regime 1)} {\hspace{-.2cm}}  & 0.26 & 0.42 & 0.38 & 0.34 & 0.34 & 0.32 & 0.39 & 0.25 & 0.49\\
$\hat \btau_2^{EM}${\scriptsize (Regime 2)}  {\hspace{-.2cm}}  & 0.54 & 0.63 & 0.64 & 0.63 & 0.59 & 0.65 & 0.61 & 0.55 & 0.66\\
\end{tabular}

\vspace{.2cm}

\begin{tabular}{c  c  c  c  c  c  c  c  c   }
\multirow{2}{*}{Tree 2}
 &
 {\footnotesize DTE,DBK$\vert$ }
 &
 {\footnotesize SIE,ALV$\vert$}
 &
 {\footnotesize SAP,DBK$\vert$
}
 &
 {\footnotesize DAI,DBK$\vert$ 
}
 &
 {\footnotesize DAI,BAS$\vert$ 
}
 & {\footnotesize SIE,BAY$\vert$ 
}
 &
  {\footnotesize EOA,SIE$\vert$
}
 &
 {\footnotesize BAS,RWE$\vert$
}\\

&  {\footnotesize ALV} & {\footnotesize DBK} & {\footnotesize SIE} & {\footnotesize SIE} & {\footnotesize SIE} & {\footnotesize BAS} & {\footnotesize BAS} & {\footnotesize EOA} \\
\hline
copula & G & G & G & G & G & G & G & N \\
$\hat \btau_1^{EM}${\scriptsize (Regime 1)} {\hspace{-.2cm}}  & 0.16 & 0.16 & 0.19 & 0.22 & 0.19 & 0.10 & 0.12  & 0.07\\
$\hat \btau_2^{EM}${\scriptsize (Regime 2)} {\hspace{-.2cm}}  & 0.22 & 0.33 & 0.13 & 0.23 & 0.20 & 0.25 & 0.25 & 0.18\\
\end{tabular}
\captionof{table}{Values of Kendall's $\tau$ determined for the DAX data set using the EM algorithm.}
\label{dax_app_table}
\end{minipage}
}

\bibliography{references}

\end{document}